\documentstyle[emulateapj,graphicx,epsfig]{article}

\def\meszaros{M\'{e}sz\'{a}ros }

\begin{document}

\title{X-ray flares: late internal and late external shocks}

\author{X. F. Wu$^{1,2,3,4}$, Z. G. Dai$^{3}$, X. Y. Wang$^{3}$, Y. F. Huang$^{3}$, L. L. Feng$^{1,2,4}$, T. Lu$^{1,2,4}$}
\affil{$^1$Purple Mountain Observatory, Chinese Academy of
Sciences, Nanjing, 210008, China; Email:xfwu@pmo.ac.cn,
fengll@pmo.ac.cn, tlu@pmo.ac.cn \\ $^2$National Astronomical
Observatories, Chinese Academy of Sciences, Beijing, 100012,
China;
\\ $^3$Department of Astronomy, Nanjing University,
Nanjing 210093, China;
\\ Email:dzg@nju.edu.cn, xywang@nju.edu.cn, hyf@nju.edu.cn
\\ $^4$Joint Center for Particle Nuclear Physics and Cosmology (CPNPC), Nanjing
210093, China;}

\begin{abstract}
We analyze several recently detected gamma-ray bursts (GRBs) with
late X-ray flares in the context of late internal shock and late
external shock models. We find that the X-ray flares in GRB 050421
and GRB 050502B originate from late internal shocks, while the
main X-ray flares in GRB 050406 and GRB 050607 may arise from late
external shocks. Under the assumption that the central engine has
two periods of activities, we get four basic types of X-ray light
curves. The classification of these types depends on which period
of activities produces the prompt gamma-ray emission (Type 1 and
Type 2: the earlier period; Type 3 and Type 4: the late period),
and on whether the late ejecta catching up with the early ejecta
happens earlier than the deceleration of the early ejecta (Type 1
and Type 3) or not (Type 2 and Type 4). We find that the X-ray
flare caused by a late external shock is a special case of Type 1.
Our analysis reveals that the X-ray light curves of GRBs 050406,
050421, and 050607 can be classified as Type 1, while the X-ray
light curve of GRB 050502B is classified as Type 2. However, the
X-ray light curve of GRB 050406 is also likely to be Type 2. We
also predict a long-lag short-lived X-ray flare caused by the
inner external shock, which forms when a low baryon-loading
long-lag late ejecta decelerates in the non-relativistic tail of
an outer external shock driven by an early ejecta.
\end{abstract}


\section{Introduction}
In the pre-{\it Swift} era, only a few early optical afterglows of
gamma-ray bursts (GRBs) were detected, while the observations of
X-ray afterglows usually started at $\sim10^{4}$ seconds after the
prompt trigger. Since early afterglows contain important
information about GRB central engines, understanding the early
afterglows is therefore one of the most interesting scientific
goals of the NASA's {\it Swift} satellite (Gehrels et al. 2004).
After about one year of operations, a significant fraction of well
localized GRBs have not been detected with early optical emissions
down to moderate limiting magnitudes, although the UV/Optical
Telescope (UVOT) on board {\it Swift} slewed to the error circle
of position quickly after the burst (Roming et al. 2005). On the
other hand, observations by the X-ray Telescope (XRT) on board
{\it Swift} have revealed several new features of X-ray emissions.
First, steep declines in some X-ray light curves at the transition
from prompt phase to afterglow phase have been discovered, which
are interpreted as tail emissions of prompt GRBs (Tagliaferri et
al. 2005; Zhang et al. 2005; Nousek et al. 2005). Secondly, a
portion of early X-ray afterglows have shallow-than-normal
temporal decays before they enter ``normal'' decaying phase
(Nousek et al. 2005; Zhang et al. 2005). Thirdly, late X-ray
flares have been observed in several GRBs (Burrows et al. 2005a;
see also Burrows et al. 2005b for a review). Piro et al. (2005)
also reported that X-ray flares were discovered in a few GRBs
(e.g., GRBs 011121 and 011211) by the Italian-Dutch {\it Beppo}SAX
satellite. GRBs are usually divided into two main classes by their
durations and spectral hardness ratios, i.e., long GRBs have
durations larger than 2 seconds and softer spectra while short
GRBs have durations shorter than 2 seconds and harder spectra
(Kouveliotou et al. 1993). Up to now, both long GRBs, including
X-ray rich bursts and a high redshift burst (GRB 050904), and
short GRBs have been found with late X-ray flares (Galli \& Piro
2005; Watson et al. 2005; Fox et al. 2005; Barthelmy et al. 2005).

One leading explanation for the early shallow decaying X-ray
afterglows is that the forward external shock has continuous
energy injection from the long active central engine (Zhang et al.
2005; Nousek et al. 2005; Dai \& Lu 1998a; Zhang \& \meszaros
2002; Dai 2004; Wei, Yan \& Fan 2005). Another leading explanation
is that the Lorentz factor of the GRB ejecta has a distribution
shaped by the central engine so that the behind slower material
catches up with the ahead faster material when the latter is
decelerated in the circum-burst medium, acting as a continuous
energy injection (Zhang et al. 2005; Nousek et al. 2005; Rees \&
\meszaros 1998; Sari \& \meszaros 2000; Zhang \& \meszaros 2002;
Granot \& Kumar 2005). In this explanation the central engine does
not need to be active for a long time (Granot \& Kumar 2005). It
is also possible that the early shallow decaying X-ray afterglows
are caused by late continuous energy releases if the GRB ejecta is
initially dominated by Poynting flux (Zhang \& Kobayashi 2004).

As for X-ray flares, both late external shock model and late
internal shock model have been proposed (Piro et al. 2005; Burrows
et al. 2005; Fan \& Wei 2005; Zhang et al. 2005). Piro et al.
(2005) and Galli \& Piro (2005) argued for the late external shock
model because they found there is no obvious spectral evolution in
some X-ray flares, while Burrows et al. (2005b) preferred the late
internal shock model based on several other X-ray flares which
showed opposite features. These two conclusions are only based on
their qualitative and empirical analysis. In this paper we
quantitatively compare the X-ray flares with the predictions of
the above two models. In Section 2 we describe the intrinsic X-ray
light curves from the internal and external shocks. Then we
analyze four GRBs with X-ray flares in Section 3 by comparing with
the ``delayed'' (relative to the GRB trigger) intrinsic light
curves of the late internal shock and late external shock. In
Section 4 we establish four basic types of X-ray light curves,
assuming that the central engine has two periods of activities. We
also classify the four GRBs by these types. Our discussion and
conclusions about X-ray flares are presented in Section 5.

\section{Intrinsic light curves of internal and external shock emissions}
Suppose that an intrinsic light curve of the internal/external
shock emissions comprises an initial rise and a subsequent decay,
which can approximately be described by the following broken
power-law function of the observer's time,
\begin{equation}
F_{\nu}=\cases{F_{\nu,pk}(t/t_{b})^{\alpha_{1}}, & $t<t_{b}$, \cr
F_{\nu,pk}(t/t_{b})^{\alpha_{2}}, & $t>t_{b}$,} \label{eq:Fnu}
\end{equation}
where $\alpha_{1}$ ($>0$), $\alpha_{2}$ ($<0$) are the temporal
indices before and after the peak time $t_{b}$, and $F_{\nu,pk}$
is the peak flux density correspondingly. The FWHM width of the
rising and falling time scale of the light curve are $\delta
t_{r}=(1-2^{-1/\alpha_{1}})t_{b}$ and $\delta
t_{f}=(2^{-1/\alpha_{2}}-1)t_{b}$, respectively.

In the internal shock model, two shells with a difference in
Lorentz factor ($\Gamma$) $\Delta\Gamma\sim\Gamma\gg1$ and a time
delay $\Delta t$ by the central engine collide at the radius
$R_{\rm{int}}\sim 2\Gamma^{2}c\Delta t$ (Paczynski \& Xu 1994).
During the interaction of these two shells, the observed emission
flux rises rapidly. The high time-resolution observations of
single-pulse GRBs (e.g. GRB 971208, Connaughton et al. 1997) and
main long pulses in usual GRBs reveal that the emission rises
linearly with time, i.e., $\alpha_{1}\sim 1$. One should keep in
mind that this value is just empirical. The falling behavior of
the light curve is mainly attributed to high-latitude emission
(also known as tail emission) of the fireball, if the width of the
fireball can be neglected (Fenimore, Madras \& Nayakshin 1996;
Kumar \& Panaitescu 2000). Let the co-moving flux density
$F_{\nu^{\prime}}^{\prime}\propto \nu^{\prime\beta}$ be uniform in
the fireball, the observed flux density is $F_{\nu}\propto
F_{\nu^{\prime}}^{\prime}{\cal{D}}^{2}(d\Sigma/dt)$, where
${\cal{D}}=1/\Gamma(1-\cos\theta)$ is the Doppler factor and the
surface area $\Sigma$ satisfies $d\Sigma=2\pi
R_{\rm{int}}^{2}\sin\theta d\theta\propto d\theta^{2}$ (Panaitescu
et al. 2005). The time delay between photons emitted at an angle
$\theta$ and those at the line of sight (LOS, $\theta=0$)is
$t-t_{b}=R_{\rm{int}}(1-\cos\theta)/c\approx
R_{\rm{int}}\theta^{2}/2c=(\theta\Gamma)^{2}\Delta t$, which
further leads to $d\Sigma/dt=$constant and
${\cal{D}}\approx2\Gamma/[(t-t_{b})/\Delta t+1]$. Since
$\nu^{\prime}=\nu/{\cal{D}}$, the theoretical falling behavior of
the internal shock emission is $F_{\nu}\propto
{\cal{D}}^{2-\beta}\propto[(t-t_{b})/\Delta t+1]^{-(2-\beta)}$
(Kumar \& Panaitescu 2000), or exactly
\begin{equation}
F_{\nu}=F_{\nu,pk}[(t-t_{b})/\Delta t+1]^{-(2-\beta)},
\phantom{ss}t_{b}<t.
\end{equation}
The initial ($t_{b}<t<\Delta t+t_{b}$) decaying part of the light
curve can be approximated by an exponential function of time,
i.e., $F_{\nu}\propto e^{-t/\tau}$, where $\tau=\Delta
t/(2-\beta)\sim \Delta t/3$. This behavior has been observed in
most of GRBs. At late times ($t\gg\Delta t+t_{b}$), the tail
emission decays as a power law function of time, $F_{\nu}\propto
t^{-(2-\beta)}$, which has been confirmed by recent X-ray
observations conducted during the prompt GRB phase by the
{\it{Swift}} satellite (Tagliaferri et al. 2005; Nousek et al.
2005). The decaying time scale (FWHM) of internal shock emission
is therefore $\delta t_{f}=[2^{1/(2-\beta)}-1]\Delta
t\sim(\ln2)\tau$, not directly related to $t_{b}$. It should be
noted that the observed $\delta t_{r}/\delta t_{f}$ is
$\sim0.3-0.5$ in prompt GRBs (Norris et al. 1996). The temporal
index $\alpha_{2}=d\ln F_{\nu}/d\ln t=(\beta-2)t/(t-t_{b}+\Delta
t)$, ranges from $-(2-\beta)t_{b}/\Delta t$ at $t\sim t_{b}$ to
$-(2-\beta)$ when $t\gg t_{b}+\Delta t$. In general, the light
curve of internal shock emission can be roughly depicted by eq.
(\ref{eq:Fnu}), especially for the rising and very late decaying
portions with $\alpha_{1}\sim 1$ and $\alpha_{2}=-(2-\beta)$.

In the external shock model, if the radial width of the GRB ejecta
is smaller than a critical value (thin shell case), the peak time
$t_{b}$ is regarded as the deceleration time $t_{\rm{dec}}$ of the
external shock sweeping into the circum-burst medium, driven by
the ejecta. On the other hand, if the width of the ejecta is
larger than that critical value (thick shell case), $t_{b}$ is the
observed crossing time of the reverse shock through the original
GRB ejecta, which is about the width divided by the speed of light
(Sari \& Piran 1999). Since the typical Lorentz factor of
electrons in the forward shock is much larger than that in the
reverse shock, the X-ray emission in most common cases is
dominated by the forward shock (Sari \& Piran 1999; see, however,
Fan \& Wei 2005 and Kobayashi et al. 2005, for opposite cases
under some extreme assumptions).

If the circum-burst environment is an interstellar medium (ISM),
the reverse shock velocity is Newtonian or relativistic depending
on whether the initial Lorentz factor $\Gamma_{i}$ is smaller or
larger than the critical value
$\Gamma_{c}\approx200E_{53}^{1/8}n_{0}^{-1/8}\Delta_{i,12}^{-3/8}$,
where $n=n_{0}$ cm$^{-3}$ is the ISM density, $E=10^{53}E_{53}$
erg and $\Delta_{i}=10^{12}\Delta_{i,12}$ cm are the
isotropic-equivalent kinetic energy and initial width of the
ejecta (Zhang, Kobayashi \& \meszaros 2003). For the Newtonian
reverse shock (NRS) case, the Lorentz factor of the forward shock
roughly equals its initial value before the reverse shock crosses
the ejecta, $\Gamma\approx\Gamma_{i}$. The minimum Lorentz factor
of shock-accelerated electrons and co-moving magnetic field in the
forward shock scale as $\gamma_{m}\propto\Gamma$ and
$B\propto\Gamma n^{1/2}$ respectively, while the cooling Lorentz
factor of electrons in the dynamical time scale is
$\gamma_{c}\propto1/t\Gamma^{3}n$ (Sari, Piran \& Narayan 1998).
The characteristic frequencies and maximum flux density of the
synchrotron radiation from the forward shock are $\nu_{m}\propto
\Gamma\gamma_{m}^{2}B\propto t^{0}$, $\nu_{c}\propto
\Gamma\gamma_{c}^{2}B\propto t^{-2}$, and $F_{\nu,\rm{max}}\propto
nR^{3}\Gamma B\propto t^{3}$ (Sari \& Piran 1999). For the
relativistic reverse shock (RRS) case, the radius and Lorentz
factor of the forward shock evolve as $R\propto t^{1/2}$ and
$\Gamma\propto t^{-1/4}$, respectively (Kobayashi 2000). The
characteristic frequencies and peak flux density in this case are
$\nu_{m}\propto t^{-1}$, $\nu_{c}\propto t^{-1}$, and
$F_{\nu,\rm{max}}\propto t$. After the reverse shock crosses the
ejecta, the forward shock emission is characterized by
$\nu_{m}\propto t^{-3/2}$, $\nu_{c}\propto t^{-1/2}$ and
$F_{\nu,\rm{max}}\propto t^{0}$ (Sari, Piran \& Narayan 1998).
Table 1 summarizes the temporal indices $\alpha_{1}$ and
$\alpha_{2}$, and the corresponding spectral index $\beta$
($F_{\nu}\propto\nu^{\beta}$) of the forward shock emission in
ISM. We can see that the flux density rises rapidly in the NRS
case, while it rises quite slowly in the RRS case. Note that in
the RRS case if the crossing times $t_{m}$ (when $\nu_{m}=\nu$)
and $t_{c}$ (when $\nu_{c}=\nu$) are both earlier than the time
when the reverse shock crosses the ejecta, the peak time $t_{b}$
should be the maximum of $t_{m}$ and $t_{c}$. In such a special
case, $\alpha_{1}=1/2$ ($t_{c}<t_{m}$) or $(3-p)/2$
($t_{c}>t_{m}$), $\alpha_{2}$ equals to $-(p-2)/2$ initially and
then to $-(3p-2)/4$. The ratio of the rising to falling times
$\delta t_{r}/\delta t_{f}$ in this case ranges from $\sim
10^{-3}$ to $\sim1$, and the spectral index changes from $-1/2$ or
$-(p-1)/2$ to $-p/2$ around $t_{b}$.

If the circum-burst environment is a stellar wind with density
$n=AR^{-2}$ ($A=3\times10^{35}A_{\ast}$ cm$^{-1}$ is the wind
parameter, Chevalier \& Li 2000; Dai \& Lu 1998b), the reverse
shock velocity is Newtonian or relativistic depending on whether
the initial Lorentz factor $\Gamma_{i}$ is smaller or larger than
the critical value
$\Gamma_{c}\approx65E_{53}^{1/4}A_{\ast}^{-1/4}\Delta_{i,12}^{-1/4}$
(Zou, Wu \& Dai 2005). However, the Lorentz factor $\Gamma$ of the
forward shock remains to be a constant before the reverse shock
crosses the ejecta, although $\Gamma\approx\Gamma_{i}$ for the NRS
case while $\Gamma\ll\Gamma_{i}$ for the RRS case (Chevalier \& Li
2000; Wu et al. 2003; Zou, Wu \& Dai 2005). Since $R\propto t$,
$n\propto t^{-2}$, $B\propto t^{-1}$, $\gamma_{m}\propto t^{0}$,
and $\gamma_{c}\propto t$, we obtain $\nu_{m}\propto t^{-1}$,
$\nu_{c}\propto t$ and $F_{\nu,\rm{max}}\propto t^{0}$. After the
reverse shock crosses the ejecta, the forward shock emission is
characterized by $\nu_{m}\propto t^{-3/2}$, $\nu_{c}\propto
t^{1/2}$ and $F_{\nu,\rm{max}}\propto t^{-1/2}$ (Chevalier \& Li
2000). Table 2 summarizes the temporal and spectral indices
$\alpha_{1}$, $\alpha_{2}$, and $\beta$ of the forward shock
emission in a stellar wind. To reproduce an initial increasing
flux density in the wind environment one must require
$\nu_{c}<\nu<\nu_{m}$. If the time when $\nu_{m}$ equals to $\nu$
is earlier than the reverse shock crosses the ejecta, the peak
time $t_{b}$ should be $t_{m}$. In this case, $\alpha_{1}=1/2$,
$\alpha_{2}$ is equal to $-(p-2)/2$ initially and then to
$-(3p-2)/4$, $\delta t_{r}/\delta t_{f}\sim 10^{-3}-1$, and the
spectral index changes from $-1/2$ to $-p/2$ at around $t_{b}$.

From the above analysis we can see that the simple internal or
external shock models can not account for the observed X-ray
flares with fast rising ($\alpha_{1}\sim10$) and rapid decay
($\alpha_{2}\sim-10$)
behavior.\footnotemark\footnotetext{\label{foot:zhang}We do not
consider the effect of magnetization of GRB ejecta on the external
shock emission, which may change the temporal indices moderately
as compared to those listed in Tables 1 and 2 (Zhang \& Kobayashi
2005). Nevertheless, this effect cannot explain the observed rapid
rising and decaying features of X-ray flares.} We study the late
internal shock (LIS) model and late external shock (LES) model for
X-ray flares in detail and make a comparison between them case by
case in particular flares in the next section.

\section{time zero point effect on the delayed internal/external shock emissions}
\begin{figure*}
\plotone{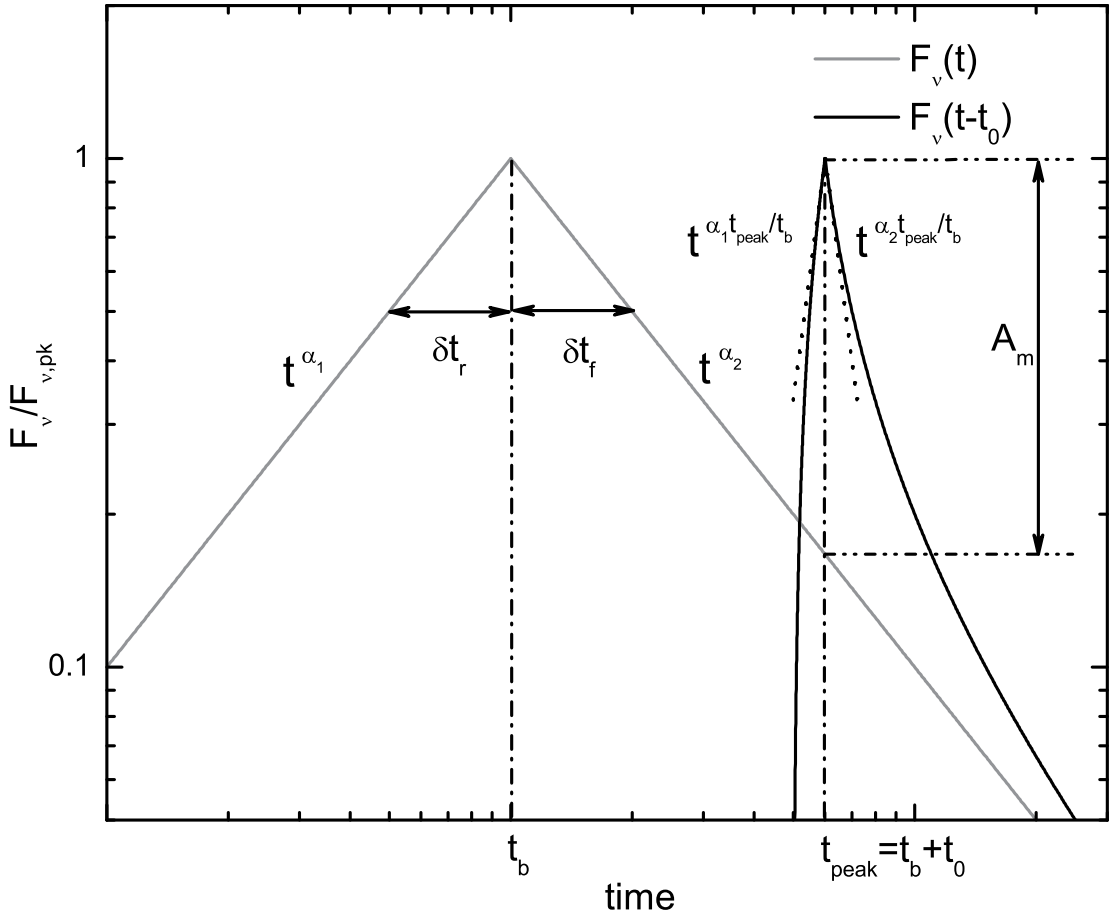} \vspace{0cm} \caption{ The light curve in grey
describes the intrinsic temporal evolution of the emissions from
one internal or external shock. The light curve in black peaked at
$t_{\rm{peak}}$ is the time-shifted intrinsic light curve by
$t_{0}$. The parameters are defined in the text.} \label{fig:t0}
\end{figure*}

Assume there exists a time delay $t_{0}$ between the trigger of a
GRB and the beginning of the above light curve as described by eq.
(\ref{eq:Fnu}). The influence of this time delay on the light
curve is known as the time zero point effect (Huang, Dai \& Lu
2002; Zhang et al. 2005). The physical reason and the conditions
for the delay within the internal and external shock models will
be presented in the next section. Here we make a quantitative
analysis of this effect. Due to the time delay, the observed light
curve plotted with the time zero point chosen as the GRB trigger
is
\begin{equation}
F_{\nu}=\cases{F_{\nu,pk}[(t-t_{0})/t_{b}]^{\alpha_{1}}, &
$t_{0}<t<t_{\rm{peak}}$, \cr
F_{\nu,pk}[(t-t_{0})/t_{b}]^{\alpha_{2}}, & $t>t_{\rm{peak}}$
\phantom{}{\rm(LES)}, \cr F_{\nu,pk}[(t-t_{\rm{peak}})/\Delta
t+1]^{-(2-\beta)}, & $t>t_{\rm{peak}}$ \phantom{}{\rm(LIS)},}
\label{eq:Fnu_t0}
\end{equation}
where the peak time is changed to $t_{\rm{peak}}=t_{0}+t_{b}$. The
observed temporal index of any segment of the light curve becomes
(e.g., Huang, Dai \& Lu 2002)
\begin{equation}
\alpha_{\rm{obs}}\equiv\frac{d\ln F_{\nu}}{d\ln
t}=\cases{\displaystyle\frac{(\beta-2)t}{t-t_{\rm{peak}}+\Delta
t}, & $t>t_{\rm{peak}}$ \phantom{}{\rm(LIS)}, \cr
\displaystyle\frac{\alpha t}{t-t_{0}}, & {\rm{otherwise}}.}
\end{equation}
Therefore, the observed temporal indices just before and after
$t_{\rm{peak}}$ are
$\alpha_{1,\rm{obs}}=\alpha_{1}t_{\rm{peak}}/t_{b}$,
$\alpha_{2,\rm{obs}}=\alpha_{2}t_{\rm{peak}}/t_{b}$ (LES), and
$\alpha_{2,\rm{obs}}=(\beta-2)t_{\rm{peak}}/\Delta t$ (LIS) (see
Fig. \ref{fig:t0}). When $t\gg t_{\rm{peak}}$, the $t_{0}$-shifted
light curve approaches to the intrinsic one. The ratio $\delta
t_{f}/t_{\rm{peak}}$, which equals
$(2^{-1/\alpha_{2}}-1)t_{b}/t_{\rm{peak}}$ in the LES model and
$(2^{1/(2-\beta)}-1)\Delta t/t_{\rm{peak}}$ in the LIS model, is
$\sim 0.1$ in all of the observed X-ray flares (Burrows et al.
2005a, b). This means that the time delay $t_{0}$ is much larger
than $t_{b}$ or $\Delta t$. A consequence of the time zero point
effect is that there exists an anti-correlation between the two
observed quantities $\alpha_{\rm{obs}}$ and $\delta
t/t_{\rm{peak}}$,
\begin{equation}
\alpha_{1,\rm{obs}}=B_{1}(\delta
t_{r}/t_{\rm{peak}})^{-1},\phantom{s}
\alpha_{2,\rm{obs}}=B_{2}(\delta t_{f}/t_{\rm{peak}})^{-1},
\label{eq:alpha12_obs}
\end{equation}
where $B_{1}=\alpha_{1}(1-2^{-1/\alpha_{1}})$ ranges from
$\sim0.4$ for $\alpha_{1}=1/2$ to $\sim 0.6$ for $\alpha_{1}=3$
and has an upper limit $B_{1}\leq0.69$, while
$B_{2}=\alpha_{2}(2^{-1/\alpha_{2}}-1)$ varies from $-1$ for
$\alpha_{2}=-1$ (typical of LES) to $-0.8$ for $\alpha_{2}=-3$
(typical of LIS, in which case $\alpha_{2}$  in eq.
[\ref{eq:alpha12_obs}] can be regarded as $\beta-2$) and has an
upper limit $B_{2}\leq-0.69$. As can be seen in Fig. \ref{fig:t0},
the ratio of the peak flux density directly measured at
$t_{\rm{peak}}$ to the one which is extrapolated from late time
($t\gg t_{\rm{peak}}$) back to $t_{\rm{peak}}$, or namely the
flare increasing factor, is
\begin{equation}
A_{m}=\cases{\Big(\displaystyle\frac{\Delta
t}{t_{\rm{peak}}}\Big)^{\beta-2}=\Big(\frac{\beta-2}{\alpha_{2,\rm{obs}}}\Big)^{\beta-2},
& {\rm{LIS}}, \cr
\Big(\displaystyle\frac{t_{b}}{t_{\rm{peak}}}\Big)^{\alpha_{2}}=\Big(\frac{\alpha_{2}}{\alpha_{2,\rm{obs}}}\Big)^{\alpha_{2}},
& {\rm{LES}}.} \label{eq:Am}
\end{equation}
This value is also roughly equal to the ratio of $F_{\nu,pk}$ to
$F_{\nu}(2t_{\rm{peak}})$, as long as $t_{0}\gg t_{b}$ or $\Delta
t$. Note that for a given $\alpha_{2,\rm{obs}}$, $A_{m}$ has a
maximum $A_{m,\rm{max}}\approx1.44^{-\alpha_{2,\rm{obs}}}$ when
$\alpha_{2}\approx0.37\alpha_{2,\rm{obs}}$. In the late internal
shock model, adopting $\beta-2\sim-3.0$ and $\delta
t_{f}/t_{\rm{peak}}\sim 0.1$, the flare increasing factor is
typically $A_{m}\sim 20$. Because the late time emission is
dominated by the external shock, the measured $A_{m,\rm{obs}}$ can
be either larger or smaller than this theoretical value, depending
on whether the backward extrapolated flux density of the external
shock emission at $t_{\rm{peak}}$ is smaller or larger than that
of the LIS emission at this time. To reduce the deviation from the
theoretical $A_{m}$ influenced by the external shock emission, the
observed $A_{m,\rm{obs}}$ may be better determined by the ratio of
$F_{\nu}(t_{\rm{peak}})$ to $F_{\nu}(2t_{\rm{peak}})$, rather than
the value extrapolated from the late time light curve. In the late
external shock model, however, the measured $A_{m}$ reflects the
true ratio. The value of $\alpha_{2}$ can be directly determined
by the post-flare light curve. Taken typical $\alpha_{2}\sim-1$
and $\delta t_{f}/t_{\rm{peak}}\sim 0.1$, we obtain $A_{m}\sim10$
for typical LES X-ray flares.

Hence, once we know $\alpha_{1}$, $\alpha_{2}$, $\delta t_{r}$,
$\delta t_{f}$, $t_{\rm{peak}}$ and $A_{m,\rm{obs}}$ of a
particular X-ray flare, the intrinsic values of $\alpha_{1}$,
$\alpha_{2}$, $t_{b}$ and $t_{0}$ are over-determined. Together
with the spectral information during the flare, we can tightly
constrain the X-ray flare models. As the X-ray spectral
information is often contained in the ratio of the count rate in
hard band to that in soft band (the so called hardness ratio
$H/S$), we convert this ratio to the spectral index $\beta$ with
Fig. 2. Below we present our case studies for X-ray flares using
available data in the literature.

{\it{GRB 050406---}} The observed quantities are
$\alpha_{1,\rm{obs}}=-\alpha_{2,\rm{obs}}=6.8$,
$t_{\rm{peak}}=213$ s, $\delta t_{r}\approx\delta
t_{f}=0.20^{+0.14}_{-0.05}t_{\rm{peak}}$, and
$A_{m,\rm{obs}}\sim6$  (Romano et al. 2005; Burrows et al. 2005).
Using eq (\ref{eq:alpha12_obs}) we obtain
$\alpha_{2}=-0.56^{+0.40}_{-0.23}$, consistent with
$\alpha_{2}\sim-0.88$ estimated by eq (\ref{eq:Am}). However, it
is impossible for the flare to have a highly symmetric light curve
with $\delta t_{r}=\delta t_{f}$ and
$\alpha_{1,\rm{obs}}=-\alpha_{2,\rm{obs}}$, unless
$\alpha_{1}=-\alpha_{2}$ approaches to infinity. If we do not take
seriously on the property of the rising behavior and adopt
$\alpha_{2}\sim-0.88$, then we get $t_{b}=28$ s and $t_{0}=185$ s.
The hardness ratio of this X-ray afterglow evolves from
$0.4\pm0.3$ initially to $1.3\pm0.3$ at the peak of the hard band
($1-10$ keV) light curve, and finally decreases to a level of
$\sim0.7$ after the peak time of the soft band ($0.1-10$ keV)
light curve. This corresponds to the spectral index $\beta$ being
$-0.7^{+0.8}_{-0.3}$, $0.0^{+0.2}_{-0.1}$, and $\sim-0.3$
respectively. The late internal shock model is disfavored in this
burst due to the fact that $\alpha_{2}=\beta-2$ is significantly
smaller than $\sim-0.88$, unless $\beta\sim1.1$ which is
impossible as indicated by the observations. Although the derived
$\alpha_{2}$ favors the late external shock model, the detected
spectral evolution is difficult to be explained in this model (see
Table \ref{tab:ISM}).

{\it{GRB 050421---}} There are two early X-ray flares residing on
a $F_{\nu}\propto t^{-3.1}$ light curve. The first stronger flare
peaks at $t_{\rm{peak}}=111\pm2$ s since the trigger, while the
second weaker flare peaks at $t=154$ s (Godet et al. 2005). The
other observed quantities of the first flare are $\delta
t_{r}\approx\delta t_{f}=0.07t_{\rm{peak}}$, and
$A_{m,\rm{obs}}\sim4$ (Godet et al. 2005). Because the superposed
$t^{-3.1}$ decaying light curve, which lasts at least 500 s since
the burst trigger, must have a different origin to the two flares,
the late external shock model for the flare can be ruled out in
this burst. For the late internal shock model, the intrinsic
$A_{m}\geq A_{m,\rm{obs}}$ requires $0.27\leq2-\beta\leq8.8$,
which can be easily satisfied. Assuming $\beta=-1.0$, the
intrinsic $A_{m}$ of the first flare is $\sim50$. Therefore the
tail emission of this flare at late times is about 12 times dimmer
than the superposed $t^{-3.1}$ emission, which may originate from
the tail of the prompt GRB. The value of $t_{b}$ of the first
flare can not be well constrained.

{\it{GRB 050502B---}} There are two X-ray flares in this burst.
Here we just discuss the first very large flare peaking at
$t_{\rm{peak}}=740$ s since the trigger. Other observed quantities
of this flare are $\alpha_{1,\rm{obs}}=-\alpha_{2,\rm{obs}}=9.5$,
and $A_{m,\rm{obs}}\sim500$. The rising and falling times satisfy
$\delta t_{r}/t_{\rm{peak}}\sim0.2$, $\delta
t_{f}/t_{\rm{peak}}\sim0.1$, and $\delta t/t_{\rm{peak}}\ll1$ for
the spike at the peak of the hard band emission (Burrows et al.
2005a, b; Falcone et al. 2005). The hardness ratio decreases
slowly from $\sim1.8$ (corresponding to $\beta\sim0.1$) when
$t\sim550$ s to $\sim0.7$ ($\beta\sim-0.4$) for $t\geq850$ s and
has a minor peak at the peak of the hard band ($1-10$ keV) light
curve of the large flare. The measured $A_{m,\rm{obs}}$ is too
large to be explained, since the theoretical maximum
$A_{m,\rm{max}}\approx33$ (when $\alpha_{2}=3.5$) is much smaller
than this value. The only way to resolve such a problem is that
the basic assumption on the flare light curve (i.e., eqs.
[\ref{eq:Fnu}] \& [\ref{eq:Fnu_t0}]) is oversimplified, as has
been directly indicated by observations (Chincarini et al. 2005).
The late external shock model is ruled out directly in this GRB
because this basic assumption is quite reasonable, and because the
pre-flare light curve can be connected with the post-flare light
curve smoothly by a single power law function of time
($F_{\nu}\propto t^{-0.8}$). The late internal shock explanation
for this flare is possible, if the flux density during the whole
decay phase is described by a pure exponential function of time,
$F_{\nu}=F_{\nu,pk}\exp[-(t-t_{\rm{peak}})/\tau]$. The flare
increasing factor in this case is
$A_{m}=F_{\nu,pk}/F_{\nu}(2t_{\rm{peak}})=\exp(t_{\rm{peak}}/\tau)$.
Combining with the observed value, we get the decay time scale
$\tau\approx0.16t_{\rm{peak}}=120$ seconds. The equivalent
temporal decaying index $\alpha_{2,\rm{obs}}=t/\tau$ at
$t=t_{\rm{peak}}$ and at $t=2t_{\rm{peak}}$ is equal to $-6.2$ and
to $-12.4$, which is consistent with the observed overall decaying
index $-9.5$. The ratio $\delta
t_{f}/t_{\rm{peak}}=(\ln2)\tau/t_{\rm{peak}}\approx0.11$ also
matches the observation. Although the exponential decay assumption
for the LIS model can explain this large flare, the physical
reason of this assumption, or equivalently, why the
$t^{-(2-\beta)}$ tail emission is not detected in this flare, has
still not been answered.

{\it{GRB 050607---}} This burst has been detected with two early
X-ray flares. The first flare is a weaker one. We focus on the
second larger flare peaking at $t_{\rm{peak}}=310$ s since the
trigger. Other observed quantities of this flare are
$\alpha_{1,\rm{obs}}\approx16$, $\alpha_{2,\rm{obs}}\approx-6.5$,
$\delta t_{r}\ll\delta t_{f}\approx0.2t_{\rm{peak}}$, and
$A_{m,\rm{obs}}\approx20$ (Pagani et al. 2005). The hardness ratio
(defined as $(H-S)/(H+S)$ for this burst, Pagani et al. 2005)
decreases slowly from $\sim0.25$ ($\beta\sim0.3$) at the beginning
of the second flare to $\sim-0.25$ ($\beta\sim-0.25$) at the end
of this flare. Using $\alpha_{2,\rm{obs}}$, $\delta
t_{f}/t_{\rm{peak}}$ and $\alpha_{1,\rm{obs}}$, we find
$\alpha_{2}\sim-0.6$, $t_{b}\approx0.09t_{\rm{peak}}=29$ s,
$t_{0}=281$ s, and $\alpha_{1}\sim1.5$. The inferred value of
$\alpha_{2}$ favors the late external shock explanation, since in
the late internal shock model $\alpha_{2}=\beta-2$ is much smaller
than $-0.6$. The inferred $\alpha_{2}$ is also consistent with the
observed temporal index $\alpha_{2}=-0.58\pm0.07$ of the
post-flare light curve before $t\sim1.2\times10^{4}$ s (Burrows et
al. 2005b). However, the measured $A_{m,\rm{obs}}$ is a little
larger than the theoretical maximum $A_{m,\rm{max}}\approx11$ when
$\alpha_{2}=-2.4$, and even larger than the derived $A_{m}=4.2$
using $\alpha_{2}\sim-0.6$ by a factor of 5. This poses a severe
crisis to the late external shock model explanation.

\section{Sequence of late activities by central engines}
The X-ray flares, happened at $\sim10^{2}$ s to $10^{4}$ s since
the trigger of prompt $\gamma$-ray emission in the burster's rest
frame, are believed to originate from prolonged activities of the
central engine. Such prolonged activity may be due to long-lived
intermittent accretion of fragmented materials (or blobs) to the
central black hole. Fragmentation may take place either in the
stellar core of the progenitor during its collapse, or in the
accretion disk surrounding the black hole due to gravitational
instability at large radii of the disk (King et al. 2005; Perna,
Armitage \& Zhang 2005). Below we discuss possible sequences of
central engine activities in the context of the LIS/LES models.

\subsection{Collision between the Early and Late Ejecta}

Let us first consider that the central engine has been active for
two periods. This is the basic picture for us to understand both
prompt GRBs and late X-ray flares. The time interval between the
beginnings of these two periods is denoted as $t_{\rm{lag}}$.
During each activity, the central engine ejects one shell whose
isotropic-equivalent kinetic energy, final mean bulk Lorentz
factor and shell width are $E_{j}$, $\Gamma_{j0}$ and
$\Delta_{j0}$, respectively. We adopt $j=e$ to denote the early
ejected shell, while $j=l$ to denote the late ejected shell. Each
shell may not be uniform and have a variable speed
$\Delta\Gamma_{j0}\sim\Gamma_{j0}$ and variability time scale
$\Delta t_{j}\ll \Delta_{j0}/c$, which are required to produce
prompt gamma-ray emission by internal shocks within the shell.

In the following, instead of discussing internal shocks within the
shells, we first study the types of the collision between the
early and late ejecta, which depend on whether this collision
happens before or after the deceleration of the early shell in the
surrounding medium.\footnotemark\footnotetext{\label{foot:reverse}
For simplicity we just consider the thin shell case for the
reverse shock propagating into the early shell. The thick shell
case could be extended in the same way.} The deceleration radius
of the early shell is
\begin{equation}
R_{{\rm{dec}},e}=\cases{1.3\times10^{17}E_{e,53}^{1/3}\Gamma_{e0,2}^{-2/3}n_{0}^{-1/3}
{\rm{cm}}, & {\rm{ISM}}, \cr
4.0\times10^{15}E_{e,53}\Gamma_{e0,2}^{-2}A_{\ast}^{-1} {\rm{cm}},
& {\rm{wind}}.} \label{eq:Rdec}
\end{equation}
Subsequently, we define the deceleration time in the observer's
frame by $t_{{\rm{dec}},e}=R_{{\rm{dec}},e}/2\Gamma_{e0}^{2}c$.
Generally, the deceleration time $t_{\rm{dec}}$ is proportional to
$E^{1/(3-k)}\Gamma_{0}^{-2(4-k)/(3-k)}$, where $k=0$ corresponds
to the ISM environment while $k=2$ corresponds to the wind
environment ($n\propto R^{-k}$). We would like to point out that,
only the prompt gamma-ray burst originates from internal shocks
within the early ejecta and the collision between the early and
late ejecta happens later than the deceleration of the early
ejecta, and then the arrival time of photons emitted from the
external shock at $R_{{\rm{dec}},e}$ driven by the early ejecta
since the GRB trigger is $t_{{\rm{dec}},e}$. Usually, there are
two types of the collision between the early and late ejecta. As
we show later, if the time interval between these two ejecta is
large enough (e.g., $t_{\rm{lag}}\sim10^{5}$ seconds), a new type
of collision exists, i.e., the late ejecta is decelerated when it
catches up and sweeps into the non-relativistic tail matter of the
forward shock driven by the early ejecta. We call this type of
collision the {\it inner external shock}. Below we give a
quantitative analysis about these three types.

{\it{(i) internal shock ---}} This happens when the late ejecta is
faster and it collides with the early ejecta before the latter is
decelerated, i.e., $\Gamma_{l0}>\Gamma_{e0}$ and
$R_{\rm{col}}<R_{{\rm{dec}},e}$. Here
\begin{equation}
R_{\rm{col}}\approx2\Gamma_{e0}^{2}c(t_{\rm{lag}}-\Delta_{e0}/c)
\end{equation}
is the radius when these two shells begin to collide with each
other, and $ct_{\rm{lag}}-\Delta_{e0}$ is the distance from the
head of the late shell and the tail of the early shell. The second
condition also reads
$t_{\rm{lag}}<t_{{\rm{dec}},e}+\Delta_{e0}/c$, which requires the
early ejecta having a very high energy $E_{e}$ or a relatively low
initial Lorentz factor $\Gamma_{e0}$. Because both ejecta are
cold, this collision is the same as typical internal shocks which
are responsible for prompt GRBs.

After the internal shock has finished, the two ejecta merge into
one shell with energy $E_{m}\approx E_{e}+E_{l}$ and bulk Lorentz
factor
$\Gamma_{m0}=\sqrt{(E_{e}+E_{l})/(E_{e}/\Gamma_{e0}^{2}+E_{l}/\Gamma_{l0}^{2})}$.
For simplicity we here neglect the energy loss taken away by
radiation. The property of the subsequent afterglow is determined
by the property of the merged shell. If $E_{e}>E_{l}$, the merged
shell is dominated by the early ejecta with $E_{m}\sim E_{e}$ and
$\Gamma_{m0}\sim\Gamma_{e0}$. The deceleration radius of the
merged shell in the circum-burst environment is
$R_{{\rm{dec}},m}\sim R_{{\rm{dec}},e}$, or equivalently, the
deceleration time is
$t_{{\rm{dec}},m}=R_{{\rm{dec}},m}/2\Gamma_{m0}^{2}c\sim
t_{{\rm{dec}},e}$. If
$E_{e}<E_{l}<(\Gamma_{l0}^{2}/\Gamma_{e0}^{2})E_{e}$, the energy
of the merged shell is $E_{m}\sim E_{l}$ while the Lorentz factor
is $\Gamma_{m0}\sim\sqrt{E_{l}/E_{e}}\Gamma_{e0}$. In this case
the deceleration radius and time are $R_{{\rm{dec}},m}\sim
R_{{\rm{dec}},e}$ and
$t_{{\rm{dec}},m}\sim(E_{e}/E_{l})t_{{\rm{dec}},e}$, respectively.
If $(\Gamma_{l0}^{2}/\Gamma_{e0}^{2})E_{e}<E_{l}$, the merged
shell is dominated by the late ejecta with $E_{m}\sim E_{l}$ and
$\Gamma_{m0}\sim\Gamma_{l0}$. The deceleration radius and time of
the merged shell are $R_{{\rm{dec}},m}\sim
R_{{\rm{dec}},l}>R_{{\rm{dec}},e}$ and $t_{{\rm{dec}},m}\sim
t_{{\rm{dec}},l}$. From the above analysis, we can see that the
afterglow may not have the same origin as the prompt $\gamma$ ray
emission, if the latter is assumed to originate from internal
shocks within the early ejecta.

{\it{(ii) refreshed shock ---}} This happens when the time lag
between the early and late ejecta by the central engine is quite
large, $t_{\rm{lag}}>t_{{\rm{dec}},e}+\Delta_{e0}/c$, or the late
ejecta has a lower Lorentz factor relative to the early ejecta
($\Gamma_{l0}\leq\Gamma_{e0}$) while the time lag is not large
($t_{\rm{lag}}\leq t_{{\rm{dec}},e}$). Therefore, the collision
between these two ejecta happens after the deceleration of the
early ejecta. The hydrodynamic evolution of the external shock
driven by the early ejecta after its deceleration follows the
Blandford-McKee solution (Blandford \& McKee 1976),
\begin{equation}
\Gamma_{e}^{2}=\Gamma_{e0}^{2}\Big(\frac{R_{e}}{R_{{\rm{dec}},e}}\Big)^{-(3-k)}.
\end{equation}
The external shock, or blast wave, can be approximated as a thin
shell with a width $\Delta R_{e}=R_{e}/2(3-k)\Gamma_{e}^{2}$. When
the late ejecta catches up with the tail of the thin shell at
$R_{\rm{col}}$, the external shock reaches
$R_{e}=R_{\rm{col}}+\Delta R_{e}$ with a Lorentz factor
$\Gamma_{e}\approx\Gamma_{e}(R_{\rm{col}})$. In the rest frame of
the central engine, the time elapsed since the first ejection by
the central engine is
$t_{\rm{elapse}}=R_{{\rm{dec}},e}/\beta_{e0}c+\int_{R_{{\rm{dec}},e}}^{R_{e}}dR/\beta_{e}c$,
where $\beta=\sqrt{1-1/\Gamma^{2}}$ is the speed in units of $c$.
This time can also be estimated by the late ejecta, i.e.,
$t_{\rm{elapse}}=R_{\rm{col}}/\beta_{l0}c+t_{\rm{lag}}$. Equaling
these two elapsed times, we obtain
\begin{equation}
\Big[\frac{7-2k}{2(3-k)(4-k)\Gamma_{e}^{2}(R_{\rm{col}})}-\frac{1}{2\Gamma_{l0}^{2}}\Big]\frac{R_{\rm{col}}}{c}=\Big(t_{\rm{lag}}-\frac{3-k}{4-k}t_{{\rm{dec}},e}\Big),
\end{equation}
or
\begin{equation}
\frac{7-2k}{3-k}\frac{t_{\rm{col}}}{t_{{\rm{dec}},e}}-\frac{\Gamma_{e0}^{2}}{\Gamma_{l0}^{2}}\Big(\frac{t_{\rm{col}}}{t_{{\rm{dec}},e}}\Big)^{1/(4-k)}=\frac{t_{\rm{lag}}}{t_{{\rm{dec}},e}}-\frac{3-k}{4-k},
\end{equation}
where
$t_{\rm{col}}=R_{\rm{col}}/2(4-k)\Gamma_{e}^{2}(R_{\rm{col}})c$ is
the observed time since the detection of first photons emitted by
the early ejecta and thus is larger than $t_{{\rm {dec}},e}$. For
$\Gamma_{l0}>\Gamma_{e0}$, since the second term on the left side
of the above equation can always be neglected, the collision
happens at
\begin{equation}
t_{\rm{col}}\simeq\frac{3-k}{7-2k}\Big(t_{\rm{lag}}-\frac{3-k}{4-k}t_{{\rm{dec}},e}\Big),
\end{equation}
which is almost independent on the ratio
$\Gamma_{l0}/\Gamma_{e0}$. This means that the time when the two
ejecta begin to collide detected in the observer's frame directly
measures the time lag $t_{\rm{lag}}$ of these two ejections by the
central engine. On the other hand, for
$\Gamma_{l0}\ll\Gamma_{e0}$, the collision happens at
\begin{equation}
t_{\rm{col}}\simeq\cases{\Big(\displaystyle\frac{3-k}{7-2k}\frac{\Gamma_{e0}^{2}}{\Gamma_{l0}^{2}}\Big)^{(4-k)/(3-k)}t_{{\rm{dec}},e},
& $t_{\rm{lag}}<t_{\rm{lag,crit}}$, \cr
\displaystyle\frac{3-k}{7-2k}\Big(t_{\rm{lag}}-\frac{3-k}{4-k}t_{{\rm{dec}},e}\Big),
& $t_{\rm{lag}}>t_{\rm{lag,crit}}$,}
\end{equation}
where the critical value of $t_{\rm{lag}}$ for a given ratio
$\Gamma_{l0}/\Gamma_{e0}$ is
$t_{\rm{lag,crit}}\simeq[(3-k)/(7-2k)]^{1/(3-k)}(\Gamma_{e0}/\Gamma_{l0})^{2(4-k)/(3-k)}t_{{\rm{dec}},e}$.
Therefore, if the late ejecta moves slower than the early ejecta,
we can use the collision time $t_{\rm{col}}$ to constrain the
contrast between the Lorentz factors of these two ejecta when
$t_{\rm{lag}}$ is small, or to directly obtain $t_{\rm{lag}}$ when
it is large.

The strength of the collision is dependent on the ratio of
$\Gamma_{l0}$ to $\Gamma_{e}(R_{\rm{col}})$. For the cases of
$\Gamma_{l0}>\Gamma_{e0}$ and of $\Gamma_{l0}<\Gamma_{e0}$ while
$t_{\rm{lag}}>t_{\rm{lag,crit}}$, we obtain
$\Gamma_{l0}/\Gamma_{e}(R_{\rm{col}})=(\Gamma_{l0}/\Gamma_{e0})[(3-k)t_{\rm{lag}}/(7-2k)t_{{\rm{dec}},e}]^{(3-k)/2(4-k)}$,
which is larger than $[(3-k)/(7-2k)]^{1/2}$. On the other hand,
for the case of $\Gamma_{l0}<\Gamma_{e0}$ while
$t_{\rm{lag}}<t_{\rm{lag,crit}}$, we get
$\Gamma_{l0}/\Gamma_{e}(R_{\rm{col}})=[(3-k)/(7-2k)]^{1/2}$.
Therefore, a violent collision requires the initial Lorentz factor
of the late ejecta much larger than that of the early ejecta, or
the time interval between these ejecta by the central engine long
enough (see also Zhang \& \meszaros 2002). A forward shock
propagating into the downstream fluid of the external shock and a
reverse shock propagating into the late ejecta are developed
during this collision. Since the total energy of the external
shock is increased after the two have merged as a whole, the
collision is also known as the refreshed shock. To produce an
obvious variability, flattening or bump in the afterglow light
curve by the refreshed shock requires that the energy in the late
ejecta must be comparable to or larger than that in the early
ejecta, or equivalently, in the initial external shock (Kumar \&
Piran 2000; Zhang \& \meszaros 2002). Because the very complicated
shock jump conditions arises when considering the forward shock
sweeping into a relativistical hot medium, the duration of the
whole collision must be calculated numerically, and it is often
found to be much longer than $t_{\rm{col}}$ in the observer's
frame. Typical signatures from refreshed shocks therefore can not
be directly used to explain X-ray flares (see figures in Zhang \&
\meszaros 2002). However, these signatures are naturally expected
to appear in afterglow light curves after X-ray flares are over
and remnants of X-ray flares as late ejecta will collide with the
ahead external shock.

{\it{(iii) inner external shock ---}} Above we take the
approximation that the gas in the external shock driven by the
early ejecta is compressed into a thin shell with a single bulk
Lorentz factor. In fact, according to Blandford \& McKee (1976),
the distributions of the Lorentz factor $\gamma_{\rm{BM}}$ and the
number density $n_{\rm{BM}}$ (in the observer's frame) of shocked
media can be described as simple functions of the self-similar
variable $\chi$,
\begin{equation}
\gamma_{\rm{BM}}=2^{-1/2}\Gamma_{e}\chi^{-1/2},n_{\rm{BM}}=2\Gamma_{e}^{2}\chi^{-(7-2k)/(4-k)}n,
\end{equation}
where $\chi=1+2(4-k)(\Delta R/R_{e})\Gamma_{e}^{2}$, $\Delta
R=R_{e}-r_{e}$ is the radial distance of the shocked media to the
shock front, and $n=AR^{-k}$ is the circum-burst media density.
Far behind the shock front, say $r_{e}\leq R_{e}/2$, a uniform
cavity or bubble is shaped with the number density,
\begin{equation}
n_{{\rm{bub}}}=\cases{3.7\times10^{-5}E_{e,53}^{-3/4}n_{0}^{7/4}R_{e,17}^{9/4}
{\rm{cm}}^{-3}, & {\rm{ISM}}, \cr
5.6\times10^{2}E_{e,53}^{-1/2}A_{\ast}^{3/2}R_{e,15}^{-3/2}
{\rm{cm}}^{-3}, & {\rm{wind}}.}
\end{equation}
Note that this bubble is cool and quasi-static. The shocked media
with $\gamma_{\rm{BM}}\simeq1$ is located at
$r_{e,n}=[(15-4k)/(16-4k)]R_{e}$. The total mass in the
non-relativistic tail of the external shock is
\begin{equation}
M_{e,n}=\int_{0}^{r_{e,n}}4\pi
r^{2}n_{\rm{BM}}m_{p}dr=C_{k}\Gamma_{e}^{2(k-3)/(4-k)}M_{\rm{sw}},
\end{equation}
where $M_{\rm{sw}}=4\pi nm_{p}R_{e}^{3}/(3-k)$ is the mass of
swept circum burst media, and the coefficient $C_{k}=0.96$ for
$k=0$ and $C_{k}=0.43$ for $k=2$.

Suppose that the late ejecta reaches $r_{e,n}$ while still not
being decelerated in the non-relativistic tail of the external
shock. In the rest frame of the central engine, the time elapsed
since the first ejection by the central engine is
$t_{\rm{elapse}}=R_{{\rm{dec}},e}/\beta_{e0}c+\int_{R_{{\rm{dec}},e}}^{R_{e}}dR/\beta_{e}c$.
It is also equal to
$t_{\rm{elapse}}=r_{e,n}/\beta_{l0}c+t_{\rm{lag}}$. Assuming
$\Gamma_{l0}$, $\Gamma_{e}$ much larger than unity, we have
$r_{e,n}\simeq(15-4k)c[t_{\rm{lag}}-(3-k)t_{{\rm{dec}},e}/(4-k)]$,
or the external shock radius
$R_{e}\simeq4(4-k)c[t_{\rm{lag}}-(3-k)t_{{\rm{dec}},e}/(4-k)]$.
Since $R_{e}>R_{{\rm{dec}},e}$, the time lag between the two
ejections is required to be very long,
\begin{equation}
t_{\rm{lag}}>\Gamma_{e0}^{2}t_{{\rm{dec}},e}/[2(4-k)].
\end{equation}
For example, adopting $\Gamma_{e0}\sim100$ and
$t_{{\rm{dec}},e}\sim100$ s, we require $t_{\rm{lag}}\geq1$ day.
This long lag challenges the most popular central engine models
for GRBs, however, it is supported by recent observations (e.g.,
Fox et al. 2005 for an X-ray flare happened at 16 days after the
short burst GRB 050709). Now we turn to derive the critical value
of $\Gamma_{l0}$, i.e., $\Gamma_{l0,\rm{crit}}$. A late ejecta
with this critical $\Gamma_{l0}$ will be decelerated exactly at
$R_{{\rm{dec}},l}=r_{e,n}$, where $R_{{\rm{dec}},l}$ is the
deceleration radius of the late ejecta. Using
$E_{l}/\Gamma_{l0,\rm{crit}}^{2}c^{2}=M_{e,n}$ and $E_{e}\approx
M_{\rm{sw}}\Gamma_{e}^{2}c^{2}$, we get
\begin{eqnarray}
\Gamma_{l0,\rm{crit}}&\approx&\Big(\frac{E_{l}}{E_{e}}\Big)^{1/2}\Gamma_{e}^{(7-2k)/(4-k)}\nonumber\\
&\approx&\Big(\frac{E_{l}}{E_{e}}\Big)^{1/2}\Gamma_{e0}^{7-2k}\Big[\frac{t_{{\rm{dec}},e}}{2(4-k)t_{\rm{lag}}}\Big]^{(3-k)(7-2k)/[2(4-k)]},
\end{eqnarray}
where $C_{k}$ is ignored for simplicity. Late ejecta with same
$t_{\rm{lag}}$ but $\Gamma_{l0}>\Gamma_{l0,\rm{crit}}$ will be
decelerated before reaching $r_{e,n}$, i.e.,
$R_{{\rm{dec}},l}<r_{e,n}$. Therefore, the external shock driven
by the late ejecta and propagating into the non-relativistic tail
of the outer external shock are called the {\it inner external
shock}. Since the mass in the bubble within $R_{{\rm{dec}},l}$ is
equal to $M_{e,n}(R_{{\rm{dec}},l}/r_{e,n})^{3}$, the deceleration
radius of the late ejecta is
$R_{{\rm{dec}},l}=r_{e,n}(E_{l}/M_{e,n}\Gamma_{l0}^{2}c^{2})^{1/3}$.
Again, letting the elapsed time
$R_{{\rm{dec}},e}/\beta_{e0}c+\int_{R_{{\rm{dec}},e}}^{R_{e}}dR/\beta_{e}c$
equal to $R_{{\rm{dec}},l}/\beta_{l0}c+t_{\rm{lag}}$, we obtain
the radius of the outer external shock when the late ejecta has
been just decelerated,
\begin{equation}
R_{e}\simeq\cases{4(4-k)ct_{\rm{lag}}\Big(\displaystyle\frac{\Gamma_{l0,\rm{crit}}}{\Gamma_{l0}}\Big)^{2(4-k)/[(3-k)(7-2k)]},
& \cr \phantom{ssssssssssss} {\rm{if}}\phantom{ss}
\Gamma_{l0,\rm{crit}}<\Gamma_{l0}<D_{k}\Gamma_{l0,\rm{crit}}, &
\cr ct_{\rm{lag}}, & \cr \phantom{ssssssssssss}
{\rm{if}}\phantom{ss} D_{k}\Gamma_{l0,\rm{crit}}<\Gamma_{l0}, &}
\end{equation}
where $D_{k}=(15-4k)^{3/2}(16-4k)^{(2k^{2}-10k+9)/(8-2k)}$ is
$\sim1.3\times10^{3}$ for $k=0$ and $\sim3.9$ for $k=2$. The
distance between the outer external shock and the late ejecta when
the latter reaches its deceleration radius is always
$R_{e}-R_{{\rm{dec}},l}\simeq ct_{\rm{lag}}$. Similar to the outer
external shock, the X-ray emission from the inner external shock
peaks at the moment of its deceleration. The time scale of the
peak in the observer's frame is $\sim
t_{{\rm{dec}},l}=R_{{\rm{dec}},l}/2\Gamma_{l0}^{2}c$, which is
much shorter than the lag time $t_{\rm{lag}}$ between the inner
and outer shocks. Therefore, the observed emission from the inner
external shock behaves as a very short spike. The inner external
shock will finally catches up with the main part of the outer
external shock. Because the gas just behind each external shock
are both hot, strong forward and reverse shocks are hardly to be
developed when they begin to collide.

In Fig. 3 we visualize the above three types of collision between
the early and late ejecta. Assuming the properties of the early
ejecta are known, the type of collision depends mainly on the
ratio of the initial Lorentz factor of the late ejecta to that of
the early ejecta, $\Gamma_{l0}/\Gamma_{e0}$, and on the time lag
$t_{\rm{lag}}$ between these two ejections by the central engine.
More details can be found in the caption of Fig. 3. Here we note
that for the refreshed shock, the relative speed of the late
ejecta in the rest frame of the outer external shock must be
supersonic, i.e., larger than $c/\sqrt{3}$. This condition is
equivalent to that the relative Lorentz factor, which is equal to
$[\Gamma_{l0}/\Gamma_{e}(R_{\rm{col}})+\Gamma_{e}(R_{\rm{col}})/\Gamma_{l0}]/2$
as long as $\Gamma_{l0}$ and $\Gamma_{e}(R_{\rm{col}})$ much
larger than unity, must be greater than $1.22$ (Zhang \& \meszaros
2002). Therefore the criterion for forming a refreshed shock is
$\Gamma_{l0}/\Gamma_{e}(R_{\rm{col}})>1.92$. We can see from Fig.
3 that if the refreshed shock is strong, the observed beginning
time of the collision directly scales to the time lag by
$t_{\rm{col}}\simeq[(3-k)/(7-2k)]t_{\rm{lag}}$. This is especially
important because only strong refreshed shocks can be easily
detected by a brightening or flaring signature in the light curve.
In the following we just consider the strong collision cases,
including the internal shock case ($\Gamma_{l0}>\Gamma_{e0}$ and
$t_{\rm{lag}}<t_{{\rm{dec}},e}$), and the strong refreshed shock
cases ($\Gamma_{l0}>\Gamma_{e0}$ and
$t_{\rm{lag}}>t_{{\rm{dec}},e}$, or $\Gamma_{l0}<\Gamma_{e0}$ and
$t_{\rm{lag}}>t_{\rm{lag,crit}}$).

\subsection{Types of X-ray light curves}

Now let us turn to discuss the entire X-ray light curve from the
early and late ejecta. Temporal variabilities in at least one of
the two ejecta are required to produce the prompt GRBs. There are
four basic types of X-ray light curves, depending on the prompt
GRB emission originating from the early ejecta or from the late
ejecta, as well as the collision between the two ejecta taking
place earlier than the deceleration of the early ejecta or not.

{\it Type 1 ---} Prompt gamma-ray burst arises from internal
shocks in the early ejecta, and the collision between the early
and late ejecta happens earlier than the deceleration of the early
ejecta, i.e., $R_{\rm{col}}<R_{{\rm{dec}},e}$. This case
corresponds to $\Gamma_{l0}>\Gamma_{e0}$, and
$t_{\rm{lag}}<t_{{\rm{dec}},e}$.

The prompt GRB is produced at the internal shock radius of the
early ejecta, $R_{{\rm{int}},e}\approx2\Gamma_{e0}^{2}c\Delta
t_{e}$. If there is also time variability in the late ejecta, then
a second burst will be produced by the internal shock within the
late ejecta at $R_{{\rm{int}},l}\approx2\Gamma_{l0}^{2}c\Delta
t_{l}$. At this moment we have the equation for the elapsed times
in the burster's frame,
$R_{{\rm{int}},e}/\beta_{e0}c+(R_{\rm{GRB}}-R_{{\rm{int}},e})/c=t_{\rm{lag}}+R_{{\rm{int}},l}/\beta_{l0}c$,
where $R_{\rm{GRB}}$ is the radius of prompt gamma-ray photons
when the internal shock of the late ejecta happens. The time
interval between the beginnings of the two bursts, or the $t_{0}$
of the second burst, is
\begin{equation}
t_{0}(l)=\frac{R_{\rm{GRB}}-R_{{\rm{int}},l}}{c}=t_{\rm{lag}}+\Delta
t_{l}-\Delta t_{e}\approx t_{\rm{lag}}.
\end{equation}
When $R_{{\rm{int}},l}>R_{{\rm{int}},e}$, the main emission of the
second burst is in the X-ray band and the internal shock of the
late ejecta produces the first X-ray flare, providing $\Delta
t_{l}\ll t_{\rm{lag}}$. Since the light curves of most of observed
X-ray flares are less structured, the late ejecta is required to
be composed of a few (two or three) mini-shells. A second X-ray
flare is unavoidable because of the collision between the early
and late ejecta. The equation for the elapsed times at this
collision is
$R_{{\rm{int}},e}/\beta_{e0}c+(R_{\rm{GRB}}-R_{{\rm{int}},e})/c=t_{\rm{lag}}+R_{{\rm{col}}}/\beta_{l0}c$,
and $R_{\rm{GRB}}$ here is the radius of prompt gamma-ray photons
when the two ejecta collide with each other. The $t_{0}$ of the
second X-ray flare relative to the trigger of the prompt GRB is
\begin{equation}
t_{0}(c)=\frac{R_{\rm{GRB}}-R_{\rm{col}}}{c}\approx\Big(1+\frac{\Gamma_{e0}^{2}}{\Gamma_{l0}^{2}}\Big)t_{\rm{lag}}-\frac{\Gamma_{e0}^{2}}{\Gamma_{l0}^{2}}T_{\rm{GRB}},
\end{equation}
where $T_{\rm{GRB}}\approx\Delta_{e0}/c$ is the duration of GRB.
After the collision, the merged shell with energy $E_{m}$ and
Lorentz factor $\Gamma_{m0}$ sweeps into the circum-burst medium
and will be decelerated at $R_{{\rm{dec}},m}$. The maximal flux of
X-ray emission of the afterglow is also produced at this radius.
By using the equation for the elapsed times at the deceleration,
$R_{{\rm{int}},e}/\beta_{e0}c+(R_{\rm{GRB}}-R_{{\rm{int}},e})/c=t_{\rm{lag}}+R_{{\rm{col}}}/\beta_{l0}c+(R_{{\rm{dec}},m}-R_{{\rm{col}}})/\beta_{m0}c$,
we have the peak time of the X-ray afterglow in the observer's
frame,
\begin{eqnarray}
t_{\rm{pk}}({\rm{AG}})&=&\frac{R_{\rm{GRB}}-R_{{\rm{dec}},m}}{c}\nonumber\\
&\approx&\Big(1-\frac{\Gamma_{e0}^{2}}{\Gamma_{m0}^{2}}+\frac{\Gamma_{e0}^{2}}{\Gamma_{l0}^{2}}\Big)t_{\rm{lag}}+t_{{\rm{dec}},m},
\end{eqnarray}
where we have assumed $\Delta_{e0}/c\ll t_{\rm{lag}}$. As
discussed in the last subsection, the property of the merged shell
is determined by the energy ratio $E_{l}/E_{e}$. If $E_{l}<E_{e}$,
since $\Gamma_{m0}\sim\Gamma_{e0}$ and $t_{{\rm{dec}},m}\sim
t_{{\rm{dec}},e}>t_{\rm{lag}}$, we have
$t_{\rm{pk}}({\rm{AG}})\approx t_{{\rm{dec}},m}$, which means that
the $t_{0}$ effect can be neglected in this case. The prompt GRB
and afterglow have the same origin. If
$E_{e}<E_{l}<(\Gamma_{l0}^{2}/\Gamma_{e0}^{2})E_{e}$, since
$\Gamma_{m0}\sim\sqrt{E_{l}/E_{e}}\Gamma_{e0}$ and
$t_{{\rm{dec}},m}\sim (E_{e}/E_{l})t_{{\rm{dec}},e}$, we have
$t_{\rm{pk}}({\rm{AG}})\approx
(1-E_{e}/E_{l})t_{\rm{lag}}+t_{{\rm{dec}},m}$. The X-ray afterglow
would behave as an X-ray flare only if
$t_{{\rm{dec}},m}\ll(1-E_{e}/E_{l})t_{\rm{lag}}$, which requires
$E_{l}/E_{e}\gg t_{{\rm{dec}},e}/t_{\rm{lag}}$ and
$\Gamma_{l0}/\Gamma_{e0}\gg\sqrt{t_{{\rm{dec}},e}/t_{\rm{lag}}}$.
Inversely, the X-ray afterglow does not suffer from the $t_{0}$
effect. If $(\Gamma_{l0}^{2}/\Gamma_{e0}^{2})E_{e}<E_{l}$, since
$\Gamma_{m0}\sim\Gamma_{l0}$ and $t_{{\rm{dec}},m}\sim
t_{{\rm{dec}},l}$, we have $t_{\rm{pk}}({\rm{AG}})\approx
t_{\rm{lag}}+t_{{\rm{dec}},m}$. The criterion for the X-ray
afterglow to be an X-ray flare is $t_{{\rm{dec}},m}\ll
t_{\rm{lag}}$, or $E_{l}/E_{e}\ll
(t_{\rm{lag}}/t_{{\rm{dec}},e})^{3-k}(\Gamma_{l0}/\Gamma_{e0})^{8-2k}$
and
$\Gamma_{l0}/\Gamma_{e0}\gg\sqrt{t_{{\rm{dec}},e}/t_{\rm{lag}}}$.
In conclusion, the late external shock model for X-ray flares
requires the energy of the late ejecta is much larger than the
energy of the early ejecta, and the ratio of initial Lorentz
factors satisfies
$\Gamma_{l0}/\Gamma_{e0}\gg\sqrt{t_{{\rm{dec}},e}/t_{\rm{lag}}}$.
This model also implicitly requires a very low initial value of
$\Gamma_{e0}$, because of $t_{{\rm{dec}},e}>t_{\rm{lag}}\sim
10^{3}-10^{4}$ seconds.

{\it Type 2 ---} Prompt gamma-ray burst arises from internal
shocks in the early ejecta, and the collision between the early
and late ejecta happens later than the deceleration of the early
ejecta, i.e., $R_{\rm{col}}>R_{{\rm{dec}},e}$. We just consider a
strong refreshed shock being developed during the collision. This
corresponds to the case of $\Gamma_{l0}>\Gamma_{e0}$,
$t_{\rm{lag}}>t_{{\rm{dec}},e}$, or the case of
$\Gamma_{l0}<\Gamma_{e0}$,
$t_{\rm{lag}}>t_{\rm{lag,crit}}>t_{{\rm{dec}},e}$.

Similar to the analysis in {\it Type 1}, the prompt gamma-ray
emission is produced by internal shocks in the early ejecta at
$R_{{\rm{int}},e}\approx2\Gamma_{e0}^{2}c\Delta t_{e}$. In the
observer's frame, the X-ray afterglow from the external shock
driven by the early ejecta will peak at
\begin{equation}
t_{\rm{pk}}({\rm{AG}})=t_{{\rm{dec}},e}.
\end{equation}
If a second burst is produced by the internal shock within the
late ejecta, then the $t_{0}$ for this burst is
\begin{equation}
t_{0}(l)\approx t_{\rm{lag}},
\end{equation}
which means the second burst happens later than the deceleration
of the early ejecta in the observer's frame. Subsequently, when
the late ejecta begins to collide with the external shock
originating from the early ejecta, the equation for the elapsed
times is
$(R_{\rm{GRB}}-R_{{\rm{int}},e})/c=(R_{{\rm{dec}},e}-R_{{\rm{int}},e})/\beta_{e0}c+\int_{R_{{\rm{dec}},e}}^{R_{e}}dR/\beta_{e}c$,
where
$R_{e}=R_{\rm{col}}[1+1/2(3-k)\Gamma_{e}^{2}(R_{\rm{col}})]$. The
$t_{0}$ of the refreshed shock emission relative to the prompt
gamma-ray emission is
\begin{equation}
t_{0}(c)\approx
\frac{7-2k}{3-k}t_{\rm{col}}+\frac{3-k}{4-k}t_{{\rm{dec}},e}=t_{\rm{lag}}.
\end{equation}
This means that the refreshed shock emission overlaps the internal
shock emission from the late ejecta in the observer's frame.

{\it Type 3 ---} Prompt gamma-ray burst arises from internal
shocks in the late ejecta, and the collision between the early and
late ejecta happens earlier than the deceleration of the early
ejecta, i.e., $R_{\rm{col}}<R_{{\rm{dec}},e}$. This corresponds to
the case of $\Gamma_{l0}>\Gamma_{e0}$,
$t_{\rm{lag}}<t_{{\rm{dec}},e}$. Note that in this case the
internal shock in the early ejecta is forbidden, otherwise this
emission would precede the emission from internal shocks in the
late ejecta by $t_{\rm{lag}}$ in the observer's frame and would
trigger the detector. In other words, there should be no time
variability in the early ejecta.

After producing the prompt GRB at
$R_{{\rm{int}},l}\approx2\Gamma_{l0}^{2}c\Delta t_{l}$, the late
ejecta will collide with the early ejecta at the radius
$R_{\rm{col}}$. The equation for the elapsed times at the
collision is
$R_{{\rm{int}},l}/\beta_{l0}c+(R_{\rm{GRB}}-R_{{\rm{int}},l})/c=R_{\rm{col}}/\beta_{l0}c$.
Therefore, the beginning time $t_{0}$ of this collision is
\begin{equation}
t_{0}(c)\approx
\frac{\Gamma_{e0}^{2}}{\Gamma_{l0}^{2}}(t_{\rm{lag}}-\Delta_{e0}/c),
\end{equation}
which is smaller than the lag $t_{\rm{lag}}$. When the collision
is over, the early and late ejecta merge into a shell with energy
$E_{m}$ and Lorentz factor $\Gamma_{m0}$. The emission of X-ray
afterglow peaks when the merged shell reaches to its deceleration
radius $R_{{\rm{dec}},m}$. The equation for the elapsed time at
the deceleration is
$R_{{\rm{int}},l}/\beta_{l0}c+(R_{\rm{GRB}}-R_{{\rm{int}},l})/c=R_{\rm{col}}/\beta_{l0}c+(R_{{\rm{dec}},m}-R_{\rm{col}})/\beta_{m0}c$.
The peak time of the X-ray afterglow in the observer's frame is
\begin{equation}
t_{\rm{pk}}({\rm{AG}})\approx
t_{{\rm{dec}},m}-\Big(\frac{\Gamma_{e0}^{2}}{\Gamma_{m0}^{2}}-\frac{\Gamma_{e0}^{2}}{\Gamma_{l0}^{2}}\Big)(t_{\rm{lag}}-\Delta_{e0}/c).
\end{equation}
Below we neglect the $\Delta_{e0}/c$ term for simplicity. If
$E_{l}<E_{e}$, since $\Gamma_{m0}\sim\Gamma_{e0}$ and
$t_{{\rm{dec}},m}\sim t_{{\rm{dec}},e}>t_{\rm{lag}}$, we have
$t_{\rm{pk}}({\rm{AG}})\approx t_{{\rm{dec}},m}-t_{\rm{lag}}\sim
t_{{\rm{dec}},m}$. If
$E_{e}<E_{l}<(\Gamma_{l0}^{2}/\Gamma_{e0}^{2})E_{e}$, since
$\Gamma_{m0}\sim\sqrt{E_{l}/E_{e}}\Gamma_{e0}$ and
$t_{{\rm{dec}},m}\sim (E_{e}/E_{l})t_{{\rm{dec}},e}$, we have
$t_{\rm{pk}}({\rm{AG}})\approx
(E_{e}/E_{l})(t_{{\rm{dec}},e}-t_{\rm{lag}})\sim
t_{{\rm{dec}},m}$. If
$(\Gamma_{l0}^{2}/\Gamma_{e0}^{2})E_{e}<E_{l}$, since
$\Gamma_{m0}\sim\Gamma_{l0}$ and $t_{{\rm{dec}},m}\sim
t_{{\rm{dec}},l}$, we have $t_{\rm{pk}}({\rm{AG}})\approx
t_{{\rm{dec}},m}$. In conclusion, the $t_{0}$ effect for the light
curve of X-ray afterglow can be neglected in this case.

{\it Type 4 ---} Prompt gamma-ray burst arises from internal
shocks in the late ejecta, and the collision between the early and
late ejecta happens later than the deceleration of the early
ejecta, i.e., $R_{\rm{col}}>R_{{\rm{dec}},e}$. We just consider a
strong refreshed shock being developed during the collision. This
corresponds to the case of $\Gamma_{l0}>\Gamma_{e0}$,
$t_{\rm{lag}}>t_{{\rm{dec}},e}$, or the case of
$\Gamma_{l0}<\Gamma_{e0}$,
$t_{\rm{lag}}>t_{\rm{lag,crit}}>t_{{\rm{dec}},e}$. As in {\it Type
3}, the internal shock in the early ejecta is forbidden.

The equation for the elapsed time when the early ejecta reaches
its deceleration radius is
$R_{{\rm{dec}},e}/\beta_{e0}c=t_{\rm{lag}}+R_{{\rm{int}},l}/\beta_{l0}c+(R_{\rm{GRB}}-R_{{\rm{int}},l})/c$.
The X-ray afterglow of the external shock driven by the early
ejecta peaks at
\begin{equation}
t_{\rm{pk}}({\rm{AG}})=t_{{\rm{dec}},e}-t_{\rm{lag}}-\Delta
t_{l}\approx t_{{\rm{dec}},e}-t_{\rm{lag}}.
\end{equation}
This means that the peak emission of X-ray afterglow happens
before the main gamma-ray burst in the observer's frame. Contrary
to the above three types, the beginning time of this kind of X-ray
afterglow relative to the prompt burst is negative,
$t_{0}=-t_{\rm{lag}}$. The detected light curve of the X-ray
afterglow behaves as
\begin{equation}
F_{\nu}=F_{\nu,pk}\Big(\frac{t+t_{\rm{lag}}}{t_{{\rm{dec}},e}}\Big)^{\alpha_{2}},
\end{equation}
where $t$ is the time in the observer's frame since the trigger of
the prompt burst, and $\alpha_{2}\sim-1$ is the intrinsic temporal
decaying index. The observed temporal index is
$\alpha_{2,\rm{obs}}=[t/(t+t_{\rm{lag}})]\alpha_{2}$, which
approaches zero when $t\ll t_{\rm{lag}}$ while approaches
$\alpha_{2}$ when $t> t_{\rm{lag}}$. Such a negative $t_{0}$ could
explain the slowly decaying early X-ray afterglows recently
discovered by the {\it Swift} satellite (Zhang et al. 2005; Nousek
et al. 2005). Subsequently, when the late ejecta begins to collide
with the external shock originating from the early ejecta, the
equation for the elapsed times is
$t_{\rm{lag}}+R_{{\rm{int}},l}/\beta_{l0}c+(R_{\rm{GRB}}-R_{{\rm{int}},l})/c=R_{{\rm{dec}},e}/\beta_{e0}c+\int_{R_{{\rm{dec}},e}}^{R_{e}}dR/\beta_{e}c$,
where
$R_{e}=R_{\rm{col}}[1+1/2(3-k)\Gamma_{e}^{2}(R_{\rm{col}})]$. The
$t_{0}$ of the refreshed shock emission relative to the prompt
gamma-ray emission is
\begin{equation}
t_{0}(c)\approx
\frac{7-2k}{3-k}t_{\rm{col}}-t_{\rm{lag}}+\frac{3-k}{4-k}t_{{\rm{dec}},e}\approx0.
\end{equation}
This means that the refreshed shock emission happens at nearly the
same time with the prompt GRB produced by the internal shocks
within the late ejecta in the observer's frame.

Fig. 4 shows the above four types of X-ray light curves. Some
caveats must be given here. First, for {\it Types 1} $\&$ {\it 2},
the internal shock within the late ejecta (thin solid lines) may
not take place if there is no temporal variability in the late
ejecta. Secondly, refreshed shocks are assumed to be strong in the
figure ({\it Types 2} $\&$ {\it 4}). Weak or mildly strong
refreshed shocks will hardly influence the X-ray light curve.
Thirdly, the temporal behavior of afterglow emission in {\it Types
1} light curve depends on the property of the merger of the two
ejecta. It may behave as an X-ray flare if the property of the
merger is determined by the late ejecta with $E_{l}>E_{e}$ and
$\Gamma_{l0}/\Gamma_{e0}\gg(t_{{\rm{dec}},e}/t_{\rm{lag}})^{1/2}$.
This corresponds to the late external shock model. Otherwise the
afterglow light curve is normal and its beginning happens when the
prompt GRB is produced.

Now we begin to classify the observed X-ray light curves by using
the above results. As discussed in Section 3, GRB 050406 has one
X-ray flare with $t_{\rm{peak}}=213$ s. The derived intrinsic
$\alpha_{2}$ is $\sim -0.56$. We note that the post-flare light
curve behaves as $F_{\nu}\propto t^{-1.58\pm0.17}$ for $t<10^{3}$
s and $F_{\nu}\propto t^{-0.50\pm0.14}$ for $t>\sim4\times10^{3}$
s (Burrows et al. 2005; Romano et al. 2005). In principle, there
are two possible explanations for this burst. In the first
explanation, the X-ray light curve of GRB 050406 belongs to the
{\it Type 1} light curve. The X-ray flare is caused by the late
external shock. The intrinsic $\alpha_{2}$ is consistent with the
temporal index of the very late afterglow light curve. In the
second explanation, the X-ray light curve belongs to the {\it Type
2} light curve. The X-ray flare is interpreted as the internal
shock emission within the late ejecta. The deceleration time of
the early ejecta satisfies $t_{{\rm{dec}},e}<100$ s and the
afterglow decays as $F_{\nu}\propto t^{-1.58}$. The collision
between the late ejecta and the external shock must be weak or
mildly strong because there is no obvious signature in the light
curve just after the X-ray flare. In this explanation, we also
need a continuous energy injection from $t\sim 4\times10^{3}$ s to
explain the flattening of the light curve.

GRB 050421 definitely belongs to the {\it Type 1} light curve,
because the two X-ray flares reside on a $F_{\nu}\propto t^{-3.1}$
emission background, which is the tail emission of the prompt
gamma-ray burst. The first X-ray flare is naturally interpreted as
from the internal shock within the late ejecta, while the second
X-ray flare is from the subsequent internal shock between the late
ejecta and the early ejecta. It is intriguing of this GRB that the
tail emission lasts for about $10^{3}$ s and the afterglow
emission has not been detected.

The X-ray light curve of GRB 050502B is classified to be the {\it
Type 2} light curve. The afterglow decays as $F_{\nu}\propto
t^{-0.8}$ with the deceleration time $t_{{\rm{dec}},e}<50$ s. The
first X-ray flare peaks at $t_{\rm{peak}}\approx740$ s. The
fluence of this X-ray flare is comparable with that of the prompt
GRB. This means the energies contained in the late ejecta and in
the early ejecta are comparable. One may expect a strong refreshed
shock signature nearly simultaneously following this X-ray flare
would be detected. However, a second X-ray flare has been
detected, but it happened at $t\sim4\times10^{4}$ s, much later
than the first X-ray flare. This implies the refreshed shock is
weak, which requires the initial Lorentz factor of the late ejecta
is smaller than that of the early ejecta,
$\Gamma_{l0}<\Gamma_{e0}$, and the time lag is no longer than the
critical value, $t_{\rm{lag}}<t_{\rm{lag,crit}}$, or
$\Gamma_{l0}/\Gamma_{e0}<(t_{{\rm{dec}},e}/t_{\rm{lag}})^{(3-k)/(8-2k)}$.
Taken $t_{{\rm{dec}},e}\sim30$ s and $t_{\rm{lag}}\sim300$ s, we
have $\Gamma_{l0}/\Gamma_{e0}<0.42$ for $k=0$ (ISM) and
$\Gamma_{l0}/\Gamma_{e0}<0.56$ for $k=2$ (wind). The second X-ray
flare happens at $t\sim0.5$ day requires another late ejecta by
the central engine to catch up with the external shock and produce
a strong refreshed shock.

GRB 050607 may belong to the {\it Type 1} light curve. The first
weaker X-ray flare can be either from the internal shock within
the late ejecta, or from the internal shock between the early and
late ejecta. The second large X-ray flare is interpreted as the
late external shock emission. However, as discussed in Section 3,
the main disadvantage of this late external shock explanation is
the theoretical flare increasing factor $A_{m}\sim4$ is much
smaller than the observed $A_{m,\rm{obs}}\sim20$.

\section{Discussion and Conclusions}
In this paper we have quantitatively analyzed late X-ray flares in
the frameworks of the late internal shock model and late external
shock model. As we have shown above, both the late internal shock
model and late external shock model require late time activities
of central engines. Some of previous works suggested the late
internal shock origin for X-ray flares (Burrows et al. 2005a;
Zhang et al. 2005; Fan \& Wei 2005), while others suggested the
late external shock origin (Piro et al. 2005; Galli \& Piro 2005).
In fact, these two kinds of late shocks can coexist within a
certain gamma-ray burst. GRB 050607 may be such a possible
candidate.

Here one caveat should be made for the late external shock model.
An X-ray flare originating from the late external shock comes to
being only if the prompt gamma-ray burst is produced by internal
shocks within the early time ejecta, and the late time ejecta
carries more energy than the early one and its initial Lorentz
factor $\Gamma_{l0}$ must be larger than the initial Lorentz
factor of the early ejecta $\Gamma_{e0}$ by a factor of
$(t_{{\rm{dec}},e}/t_{\rm{lag}})^{1/2}>1$, where $t_{\rm{lag}}$ is
the time separation between the two ejecta by the central engine
and $t_{{\rm{dec}},e}$ is the deceleration time of the early
ejecta in the circum-burst medium. The merger of the two ejecta is
mainly dominated by the late ejecta with a much shorter
deceleration time relative to $t_{{\rm{dec}},e}$. Therefore the
beginning of the afterglow from the merger happens much later than
the prompt GRB and the peak of afterglow behaves as an X-ray
flare. Note this description of the late internal shock model is
different from that in Piro et al. (2005) and Galli \& Piro
(2005). They attributed the late external shock to arise from a
thick shell which is ejected by the long active central engine.
Despite different descriptions, this model requires the $t_{0}$
effect in the afterglow emission. Given $t_{0}\sim t_{\rm{lag}}$,
the smaller the deceleration time of the merger
$t_{{\rm{dec}},m}$, the more remarkable the afterglow as an X-ray
flare. Since $t_{{\rm{dec}},m}\propto
E_{m}^{1/(3-k)}\Gamma_{m0}^{-2(4-k)/(3-k)}$, a large energy or
especially a large Lorentz factor will easily cause an X-ray
flare. This result has also been obtained by Galli \& Piro (2005).

Theoretically, there are four basic types of X-ray light curves if
the central engine has two periods of activities before it
entirely ceases. According to our analysis, the observed X-ray
light curves of GRBs with X-ray flares all belong to {\it Type 1}
or {\it Type 2}. This means the prompt gamma-ray emissions of
these GRBs result from the early ejecta, while the late X-ray
flares are produced by the internal shock within the late ejecta,
by the internal shock between the early and late ejecta, or by the
late external shock. The difference between {\it Type 1} and {\it
Type 2} is that for the former the collision between the early and
late ejecta happens before the deceleration of the early ejecta,
while for the latter the situation is inverse. One may wonder
where are {\it Type 3} and {\it Type 4} X-ray light curves? In
these two types, the prompt gamma-ray burst is produced from the
late ejecta. There would be no internal shocks within the early
ejecta, or these internal shock emission is too weak to be
detected. It is also possible that the internal shock emission
within the early ejecta is very weak and regarded as the precursor
of the main burst. According to our classification, {\it Type 3}
X-ray light curves correspond to the case that the collision
between the early and late ejecta takes place before the
deceleration of the early ejecta. The emission from the collision
has a small time lag relative to the prompt gamma-ray emission and
can be regarded as the last pulse of the prompt GRB. The afterglow
emission in {\it Type 3} hardly suffers the $t_{0}$ effect.
Therefore it is hard to distinguish {\it Type 3} X-ray light
curves from the light curves in which the central engine has only
one active period. However, {\it Type 4} X-ray light curves may
have already been detected. Some of recently discovered X-ray
afterglows whose initial decay is extraordinarily slow may
originate from this type.

In this paper we also investigate a new kind of collision between
the early and late ejecta, the {\it inner external shock}. Such a
shock will be developed when the late ejecta is decelerated by the
non-relativistic tail of the outer external shock, which is driven
by the early ejecta. To develop the inner external shock requires
the time lag $t_{\rm{lag}}$ between these two ejecta is very long,
typically $\geq1$ day, and the baryon loading of the late ejecta
is very low, or equivalently, its initial Lorentz factor must be
larger than a critical value. Since the required $t_{\rm{lag}}$ is
not very extreme compared to recent discoveries and the low baryon
loading is also plausible, the inner external shock maybe exist.
Because the time scale of emission from this shock in the
observer's frame is equal to its deceleration time and therefore
is much shorter than $t_{\rm{lag}}$, the emission behaves as a
short spike. Detecting such a spike in the typical afterglow time
scale ($t\sim t_{\rm{lag}}\sim1$ day) is quite difficult. We
therefore just make a prediction of the inner external shock
emission in this paper, and have not considered such kind of
emission in the above four basic X-ray light curves.

For simplicity we have only considered the central engine having
two periods of activities. In reality the central engine may have
more than two periods of activities. The multiple X-ray flares
detected in the high redshift gamma ray burst GRB 050904 indicate
that the central engine of this burst has been active for many
periods (Watson et al. 2005; Cusumano et al. 2005). Cusumano et
al. (2005) reported that the hardness ratio for $t>100$ s is
nearly a constant, $H(14-73{\rm{ keV}})/S(1.4-14{\rm{
keV}})\sim0.2$, which corresponds to $\beta\sim-0.9$. The X-ray
light curves from multiple ejections by the central engine are
complicated and difficult to be classified. As for GRB 050904,
Zou, Xu \& Dai (2005) found that ordered late internal shocks
formed by collisions between late ejecta and the earliest ejecta
can reproduce the temporal evolution of peak luminosity of X-ray
flares and the entire X-ray spectral evolution.

At last we would like to point out other ways besides those
developed in this paper (see Section 3) to distinguish an X-ray
flare caused by a late external shock from other X-ray flares in
the same GRB by late internal shock(s). As we know in most pulses
of GRBs it is found that higher energy photons arrive to the
observer earlier than lower energy photons (e.g., Chen et al.
2005). This is the spectral-lag phenomenon. Fenimore et al. (1995)
found that the width $W$ of peaks in prompt GRBs tends to be
narrower at higher photon energy $E$, i.e., $W\propto E^{-0.4}$.
This is called the spectral-width relationship. If these two
phenomena have been detected in a particular X-ray flare, then the
X-ray flare can be proved to originate from a late internal shock.
Statistics help us to understand X-ray flares. In prompt GRBs,
several relationships have been found, such as the isotropic
gamma-ray released energy $E_{\rm{iso}}$ and spectral peak photon
energy $E_{p}$ relationship (Amati et al. 2002), the peak
luminosity $L_{p}$ and spectral peak photon energy $E_{p}$
relationship (Yonetoku et al. 2004), the peak luminosity $L_{p}$
and spectral time lag $\tau$ relationship (Norris, Marani \&
Bonnell 2000). Moreover, Liang, Dai \& Wu (2004) found there
exists a relationship between the flux and $E_{p}$ within a GRB
with time resolved spectra (see also Borgonovo \& Ryde 2001).
Therefore, investigating the above relations in X-ray flares is
urgent since it can provide direct evidence that whether late
X-ray flares and prompt GRBs have the same mechanism.

 {\acknowledgments
This work was supported by the National Natural Science Foundation
of China (grants 10573036, 10503012, 10473023, 10403002, 10233010,
and 10221001), the Special Funds for Major State Basic Research
Projects, and the Foundation for the Authors of National Excellent
Doctoral Dissertations of China (project 200125). }

\begin{deluxetable}{ccccccc}
\normalsize \tablecolumns{6} \tablewidth{0pc}
\tablecaption{Temporal and spectral indices ($F_{\nu}\propto
t^{\alpha}\nu^{\beta}$) of the forward shock emission in
ism\label{tab:ISM}} \tablehead { \colhead {} & \colhead {} &
\colhead {$t<t_{b}$ (NRS)} & \colhead
{$t<t_{b}$ (RRS)} & \colhead {($t>t_{b}$)} & \colhead {$\delta t_{r}/\delta t_{f}$}\\
\colhead {regime} & \colhead {$\beta$} & \colhead {$\alpha_{1}$} &
\colhead {$\alpha_{1}$} & \colhead {$\alpha_{2}$} & \colhead {(NRS
vs RRS)}} \startdata
$\nu<\nu_{c}<\nu_{m}$ & $1/3$ & $11/3$   & $4/3$  & $1/6$ & none\\
$\nu<\nu_{m}<\nu_{c}$ & $1/3$ & $3$   & $4/3$  & $1/2$ & none\\
$\nu_{c}<\nu<\nu_{m}$ & $-1/2$ & $2$   & $1/2$  & $-1/4$ & $0.02$ vs $0.05$\\
$\nu_{m}<\nu<\nu_{c}$ & $-(p-1)/2$ & $3$ & $(3-p)/2$   & $-3(p-1)/4$ & $0.18$ vs $0.71$\\
$\max\{\nu_{m},\nu_{c}\}<\nu$ & $-p/2$  & $2$ & $-(p-2)/2$ & $-(3p-2)/4$ & $0.35$ vs none\\
\enddata
\tablecomments{The indices are derived under the assumptions that
the hydrodynamics is adiabatic and the emission is from the
synchrotron radiation of shock-accelerated electrons, whose
initial energy distribution follows $dN/d\gamma_{e}\propto
\gamma_{e}^{-p}$. The ratio $\delta t_{r}/\delta t_{f}$ is
estimated by assuming the observing band in the same spectral
regime around $t_{b}$ and adopting $p=2.2$.}
\end{deluxetable}

\begin{deluxetable}{cccccc}
\normalsize \tablecolumns{5} \tablewidth{0pc}
\tablecaption{Temporal and spectral indices ($F_{\nu}\propto
t^{\alpha}\nu^{\beta}$) of the forward shock emission in stellar
wind\label{tab:wind}} \tablehead { \colhead {regime} & \colhead
{$\beta$} & \colhead {$\alpha_{1}$ ($t<t_{b}$)} & \colhead
{$\alpha_{2}$ ($t>t_{b}$)} & \colhead {$\delta t_{r}/\delta
t_{f}$}} \startdata
$\nu_{c}<\nu<\nu_{m}$ & $-1/2$ & $1/2$  & $-1/4$ & $0.05$\\
$\nu_{m}<\nu<\nu_{c}$ & $-(p-1)/2$ & $-(p-1)/2$  & $-(3p-1)/4$ & none\\
$\max\{\nu_{m},\nu_{c}\}<\nu$ & $-p/2$  & $-(p-2)/2$ & $-(3p-2)/4$ & none\\
\enddata
\tablecomments{Similar to Table 1 except for a stellar wind
environment. The theoretical $\alpha_{1}$ in the case of
$\nu_{m}<\nu<\nu_{c}$ is also presented in the table, although
$\nu_{c}$ is usually below $\nu_{m}$ and the X-ray frequency for
typical parameter values. }
\end{deluxetable}

\begin{figure*}
\plotone{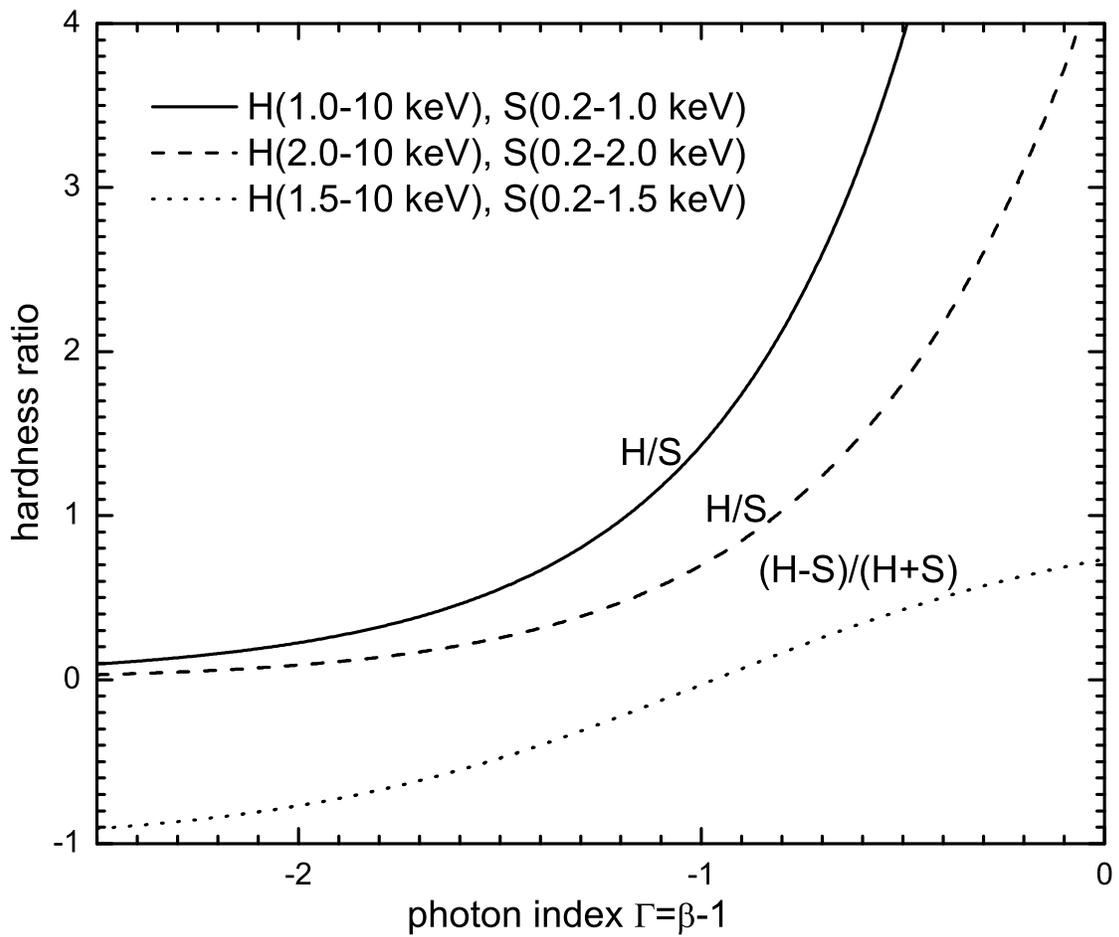} \vspace{0cm} \caption{ Hardness ratio (ratio of
the hard-band count rate to soft-band count rate) as a function of
the effective photon index $\Gamma$, assuming the $0.2-10$ keV
spectrum can be fitted by one single power-law $F_{\nu}\propto
\nu^{\beta}$.} \label{fig:hardness_ratio}
\end{figure*}

\begin{figure*}[htb]
  \begin{center}
  \centerline{ \hbox{
  \epsfig{figure=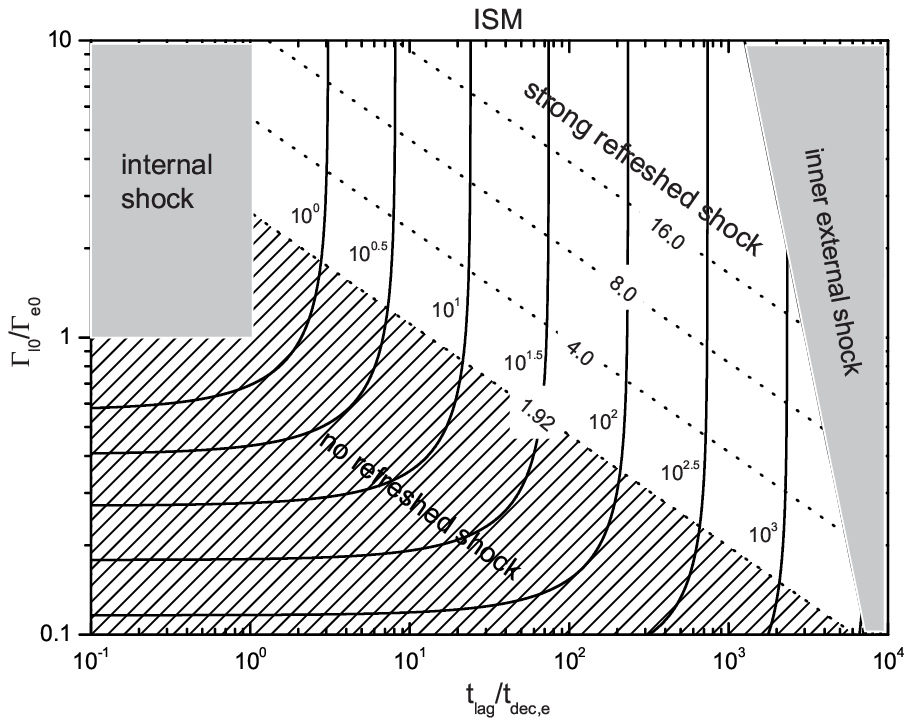,width=2.2in,height=2.2in,
  bbllx=120pt, bblly=20pt, bburx=280pt, bbury=240pt}
  \hspace{.2in}
  \epsfig{figure=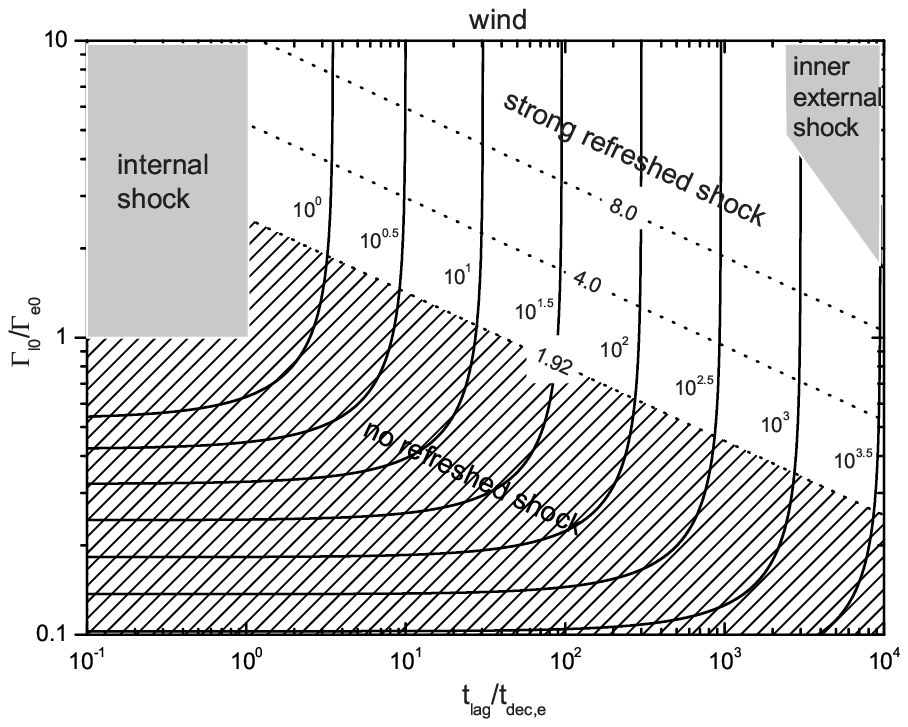,width=2.2in,height=2.2in,
  bbllx=40pt, bblly=20pt, bburx=200pt, bbury=240pt}
  }}
\caption{ Three types of collision when the late ejecta catches up
with the early ejecta, depending on the ratio of $\Gamma_{l0}$ to
$\Gamma_{e0}$ and $t_{\rm{lag}}$ (which is scaled to the
deceleration time of the early ejecta). Left and right panels
correspond to two kinds of circum burst environment, i.e., the
interstellar medium (left) and the free wind (right). The upper
left grey region in each figure denotes the region for internal
shocks. We simply neglect the width $\Delta_{e0}$ of the early
ejecta. The right grey region (plotted assuming $\Gamma_{e0}=100$,
$E_{l}=0.1E_{e}$) denotes the inner external shock region, in
which the late ejecta can be decelerated in the non-relativistic
tail of the early external shock. The rest region is for refreshed
shocks. Different solid lines correspond to different values of
$t_{\rm{col}}/t_{{\rm{dec}},e}$, ranging from $1$ to $\sim10^{4}$
as marked beside each line. Different dotted lines correspond to
different values of $\Gamma_{l0}/\Gamma_{e}(R_{\rm{col}})$, which
are marked on the lines. If
$\Gamma_{l0}/\Gamma_{e}(R_{\rm{col}})<1.92$, the refreshed shock
is suppressed by the thermal energy in the outer external shock. }
  \end{center}
  \label{fig:shocktypes}
\end{figure*}

\begin{figure*}[htb]
  \begin{center}
  \centerline{ \hbox{
  \epsfig{figure=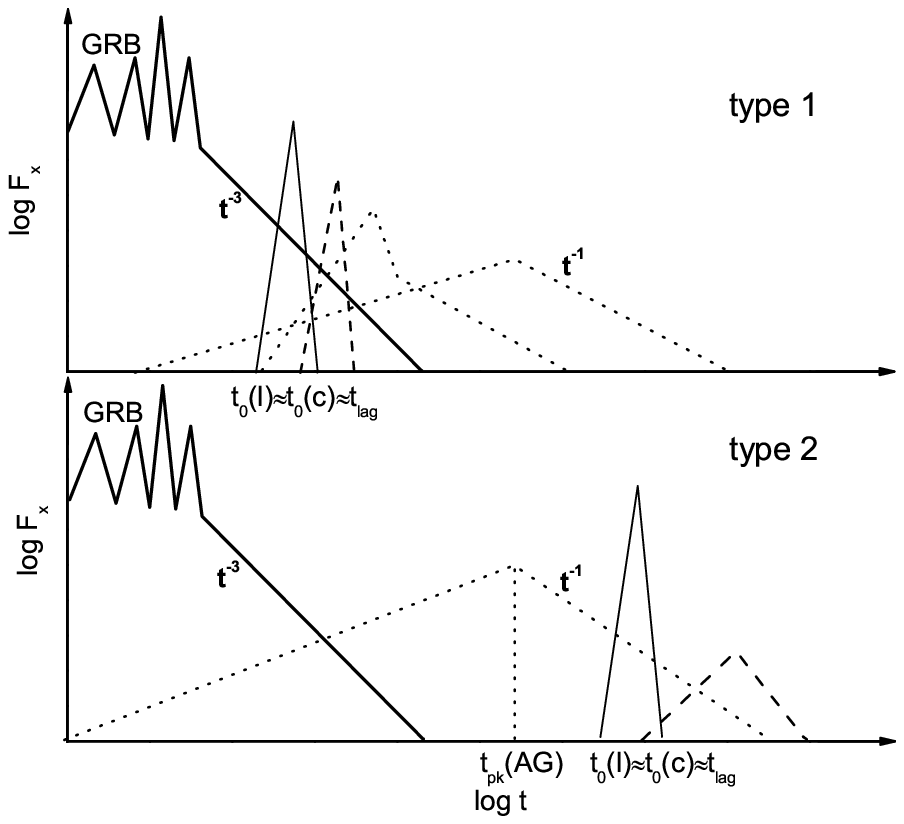,width=2.2in,height=2.2in,
  bbllx=120pt, bblly=20pt, bburx=280pt, bbury=240pt}
  \hspace{.2in}
  \epsfig{figure=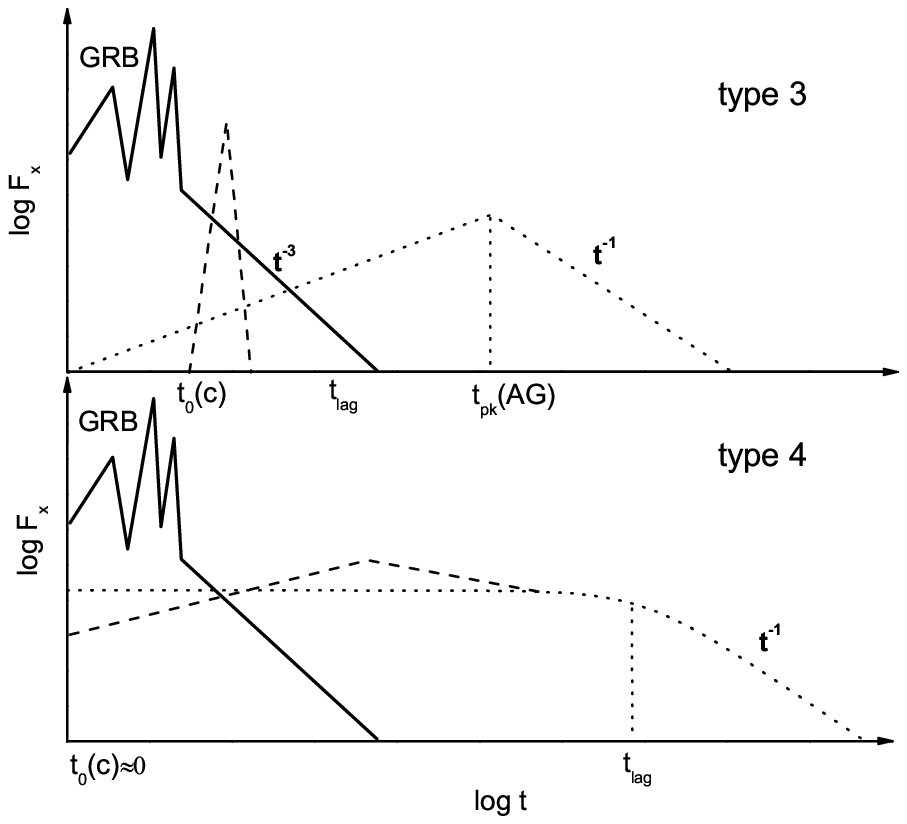,width=2.2in,height=2.2in,
  bbllx=40pt, bblly=20pt, bburx=200pt, bbury=240pt}
  }}
\caption{ Four types of X-ray light curves, depending on the
prompt GRB produced by internal shocks within the early ejecta
({\it Types 1} $\&$ {\it 2}) or by internals shocks within the
late ejecta ({\it Types 3} $\&$ {\it 4}). In each case, the X-ray
light curve is further determined by whether the collision between
the two ejecta happens earlier than the deceleration of the early
ejecta ({\it Types 1} $\&$ {\it 3}) or not ({\it Types 2} $\&$
{\it 4}). Thick solid lines represent the X-ray emission during
the prompt GRB phase. The decaying index of the tail emission of
the prompt GRB is $-2+\beta\sim-3$. Thin solid lines correspond to
the X-ray emission produced by the internal shock of the late
ejecta. Dashed lines correspond to the emission from the collision
between the two ejecta, which can be either the direct internal
shock ({\it Types 1} $\&$ {\it 3}) or the refreshed shock ({\it
Types 2} $\&$ {\it 4}). Dotted lines correspond to the emission
from the external shock.}
  \end{center}
  \label{fig:Xlightcurves}
\end{figure*}

\end{document}